
%
%
\magnification=1000
\message{-----------------------------------------------------------}
\message{
--------------- For correct dvi.file --------------------------------
}
\message{
------------------You must use----------------------------------------
}
\message{
---------------phyzzx.tex   update/debug: March 1, 1987 version-------
}
\message{
-----------------------------------------------------------------------
}
\input phyzzx
\let\refmark=\NPrefmark

\baselineskip 16pt
\hfill NIIG-DP-94-1 \par
\hfill KANAZAWA-94-03 \par
\hfill February 1994
\vskip 1.5cm
\centerline{\fourteenbf New insight into BRST anomalies in
superstring theory}
\vskip 1.0cm
\centerline{T. Fujiwara$^1$\footnote{}{$^1$ E-mail:
tfjiwa@tansei.cc.u-tokyo.ac.jp},
 Y. Igarashi$^{\rm a,2}$\footnote{}{$^2$ E-mail:
igarashi@ed.niigata-u.ac.jp}
, M. Koseki$^{\rm b,3}$\footnote{}{$^3$ E-mail: koseki@niigt.kek.jp},
R. Kuriki and T. Tabei$^{\rm c,4}$\footnote{}{$^4$ E-mail:
tabei@hep.s.kanazawa-u.ac.jp}}

\address{Department of Physics, Ibaraki University, Mito 310, Japan
\break
$^{\rm a}$ Faculty of General Education,
Niigata University,
Niigata 950-21, Japan
\break
$^{\rm b}$ Graduate School of Science and Technology,
Niigata University,
\break
Niigata 950-21, Japan
\break
$^{\rm c}$ Graduate School of Science and Technology,
Kanazawa University,
\break
Kanazawa 920-11, Japan}

\vfill
\abstract{
Based on the extended BRST formalism of Batalin, Fradkin and
Vilkovisky, we perform a general algebraic analysis of the BRST
anomalies in superstring theory of Neveu-Schwarz-Ramond.
Consistency conditions on the BRST anomalies are completely solved.
The genuine super-Virasoro anomaly is identified with the essentially
unique solution to the consistency condition without any reference to
a particular gauge for the 2D supergravity fields.
In a configuration
space where metric and gravitino fields are properly constructed,
general form of the super-Weyl anomaly is obtained from
the super-Virasoro anomaly as its descendant.
  We give a novel local action of super-Liouville type,
which plays a role
of Wess-Zumino-Witten term shifting the super-Virasoro anomaly
into the super-Weyl anomaly.
These results reveal a hierarchial relationship in
the BRST anomalies.
}
\eject
%
%
\def\Rap{Ann. Phys. (N.Y.) }
\def\Rmpl{Mod. Phys. Lett. }
\def\Rnp{Nucl. Phys. }
\def\Rpl{Phys. Lett. }
\def\Rplc{Phys. Rep. }
\def\Rpr{Phys. Rev. }
\def\Rprl{Phys. Rev. Lett. }
\def\Rptp{Prog. Theor. Phys. }
\def\Rzsp{Z. Phys.}
\def\FriA{D. Friedan, {\it in} \  ``Les Houches Summer School, 1982",
(J. B. Zuber and R. Stora, Eds.), North Holland Amsterdam, 1984.}
\def\PZa{A. M. Polyakov and A. B. Zamolodochikov, \Rmpl
{\bf A3}(1988)1213}
\def\AAZa{E. Abdalla, M. C. B. Abdalla and A. Zadra, \Rmpl
{\bf A4}(1989)849}
\def\GGMTa{S. J. Gates, M. T. Grisaru and P. K. Townsend, \Rnp
{\bf A286}(1987)1}
\def\IMNUa{M. It\=o, T. Morozumi, D. Nojiri and S. Uehara,
\Rptp {\bf 75}(1986)934}
\def\HNUa{M. Hayashi, S. Nojiri and S. Uehara, \Rzsp {\bf C31}(1986)561}
\def\TanA{Y. Tanii, \Rnp {\bf B259}(1985)677}
\def\BMGa{R. Brooks, F. Muhammad and S. J. Gates, \Rnp
{\bf B268}(1986)599}
\def\BZa{W. A. Bardeen and B. Zumino, \Rnp {\bf B244}(1984)421}
\def\GXa{M. T. Grisaru and R. -M. Xu, \Rpl {\bf B205}(1988)486}
\def\BVA{I. A. Batalin and G. A. Vilkovisky, \Rpl {\bf B69}(1977)309}
\def\LOGW{J. Louis, B.A. Ovrut, R. Garreis and J. Wess,
\Rpl {\bf B199} (1987) 57}
\def\EHLC{M. Eber, R. Heid and G. Lopes Cardoso, \Rzsp {\bf C 37} (1987) 85}
\def\Degr{M. De Groot, \Rpl {\bf B 211} (1988) 63}
\def\DegrMan{M. De Groot and P. Mansfield, \Rpl {\bf B 202} (1988) 519}
\def\Ohta{Ohta, \Rpr {\bf D 33} (1986) 1681;\Rpl {\bf B 179} (1986) 347}
\def\Schwa{J. H. Scwarz, Supple. Prog. Theor. Phys. {\bf 86} (1986) 70}
\def\BFREV{I. A. Batalin and E. S. Fradkin,
Ann. Inst. Henri Poincar\' e, {\bf 49} (1989)145}
\def\GLO{R. Garreis, J. Louis and B.A. Ovrut, \Rnp {\bf B306} (1988) 567}
\def\StrB{R. Stora, in {\it New developments in quantum field theory
 and statistical mecanics} Carg\`ese 1976  M. L\'evy, P. Mitter (eds.),
 New York, London: Plenum Press}
\def\Fuk{T. Fukai, \Rnp {\bf B299}(1988)346}
\def\SieA{W. Siegel, \Rnp {\bf B238}(1984)307}
\def\AGGA{L. Alvarez-Gaum\'e and P. Ginsparg, \Rnp {\bf B243}(1984)449}
\def\AGWa{L. Alvarez-Gaum\'e and E. Witten, \Rnp {\bf B234}(1984)269}
\def\BDHa{L. Brink, P. Di Vecchia and P. Howe,
\Rpl {\bf B65}(1976)471}
\def\CTa{T. L. Curtright and C. B. Thorn, \Rprl {\bf 48}(1982)1309}
\def\DZa{S. Deser and B. Zumino, \Rpl {\bf B65}(1976)369}
\def\FIKB{T. Fujiwara, Y. Igarashi and J. Kubo, \Rnp {\bf B358}(1991)195}
\def\FIKMa{T. Fujiwara, Y. Igarashi, J. Kubo and K. Maeda,
\Rnp {\bf B391}(1993)211}
\def\FIKMTa{T. Fujiwara, T. Tabei, Y. Igarashi, K. Madeda and J. Kubo,
\Rmpl {\bf A8}(1993)2147}
\def\FIKTb{T. Fujiwara, Y. Igarashi, J. Kubo and T. Tabei,
\Rpr {\bf D48}(1993)1736}
\def\FujB{K. Fujikawa, \Rpr {\bf D25}(1982)2584}
\def\FujC{K. Fujikawa, \Rpl {\bf B188}(1987)115}
\def\FujG{K. Fujikawa, \Rnp {\bf B291}(1987)583}
\def\FVA{E. S. Fradkin and G. A. Vilkovisky, \Rpl {\bf B55}(1975)224}
\def\GSW{M. B. Green, J. H. Schwarz and E. Witten, {\it Superstring
Theory} 1, (Cambridge University Press, Cambridge, England, 1987).}
\def\HenA{M. Henneaux, \Rplc {\bf 126}(1985)1}
\def\HwaB{D. S. Hwang, \Rpr {\bf D28}(1983)2614}
\def\KOa{M. Kato and K. Ogawa, \Rnp {\bf B212}(1983)443}
\def\KPZ{V. G. Knizhnik, A. M. Polyakov and A. B. Zamolodchikov,
\Rmpl {\bf A3}(1988)819}
\def\ManB{P. Mansfield, \Rap {\bf 180}(1987)330}
\def\PolA{A. M. Polyakov, \Rpl {\bf B103}(1981)207;211}
\def\PolB{A. M. Polyakov, \Rmpl {\bf A2}(1987)893}
\def\WZA{J. Wess and B. Zumino, \Rpl {\bf B37}(1971)95}
\def\ZumA{B. Zumino, in {\it Relativity, groups and topology} II,
Les Houches 1983, eds. B. S. DeWitt and R. Stora (North-Holland, 1984)}
\def\ZWZ{B. Zumino, Y. -S. Wu and A. Zee, \Rnp {\bf B293}(1984)477}
\def\AbeNaka{M. Abe and N. Nakanishi, \Rmpl {\bf A7}(1992)1799}
%
%
               %
               %
%
%
\def\a{\alpha}     \def\b{\beta}         \def\g{\gamma}   \def\d{\delta}
\def\e{\epsilon}     \def\z{\zeta}    
             \def\l{\lambda}
\def\m{\mu}        \def\n{\nu}           \def\x{\xi}      \def\p{\pi}
\def\r{\rho}       \def\s{\sigma}        \def\t{\tau}     \def\f{\phi}
\def\vp{\varphi}   \def\v{\psi}          \def\w{\omega}   \def\h{\eta}

\def\G{\Gamma}                \def\L{\Lambda}
\def\X{\Xi}                     \def\F{\Phi}
\def\V{\Psi}       \def\W{\Omega}        
%
%

%
%
\def\dl{\partial}   \def\7{\bigtriangledown}    \def\3{\bigtriangleup}
%
%
  
\def\ssW{{\scriptscriptstyle W}}
%
%

%

\chapter{\bf Introduction}
The Virasoro anomaly, the Weyl anomaly and the nonvanishing
square of BRST charge are known to represent a physically
equivalent obstruction in quantization of relativistic strings
at subcritical dimensions.
In spite of some
efforts using the BJL-limit \Ref\rFujBJL{\FujC\hfill\break\LOGW} and
 cohomological techniques\Ref\rManB{\ManB\hfill\break\EHLC},
however, their internal relationships have been only partially
revealed so far. This is mainly because that the considerations
on these anomalies including those in the classic references
\REFS\rPolA{\PolA}\REFSCON\rFuj{\FujB}\REFSCON\rKOa{\KOa\hfill\break\HwaB}
\REFSCON\rAGWa{\AGWa}\REFS\rFriA{\FriA}[\rPolA-\rFriA] have been
done only in
particular classes of gauges.
A general discussion where the gauge
dependence of these anomalies is satisfyingly explored is highly
desired. This goal is achieved recently in a previous
paper\Ref\rFIKMa{\FIKMa}, hereafter referred to as I, by
extensively using the generalized hamiltonian formalism of
Batalin, Fradkin and Vilkovisky (BFV)\Ref\rFVA{\FVA\hfill\break\BVA}
\foot{See \Ref\rBFVREV{\HenA\hfill\break\BFREV}
for review articles and references therein.}.
It shows up a hierarchical relationship among the anomalies, which
can not be clearly seen in the conventional gauge-fixed analyses.

The purpose of this paper is to extend the investigation given
in I for bosonic string to
superstring of Neveu-Schwarz-Ramond\foot{See \Ref\rGSW{\GSW}
for a review and references therein.}
formulated as 2D supergravity (2D SUGRA) theory coupled
with superconformal
matter\Ref\rDZa{\DZa\hfill\break\BDHa}.
This extension contains a
nontrivial task to find a suitable BRST quantization scheme for
2D SUGRA. The BRST quantization of
2D SUGRA based on the configuration space introduces a lot of
redundant variables to maintain off-shell nilpotency of the BRST
charge\REFS\rOhta{\Ohta\hfill\break\Schwa}
\REFSCON\rIMNUa{\IMNUa}[\rOhta,\rIMNUa].
The general scheme given in \Ref\rHNUa{\HNUa} includes not only
the auxiliary field
in the
supermultiplet but also some pure gauge fields associated with
local symmetries. The vanishing-curvature conditions for the
gauge fields, however, can not always be solved explicitly, and
it would make the general investigation we wish to develop rather
complicated.\foot{Since we will not assume that supersymmetry is
intact upon quantization, superspace formulations\Ref\rBMGa{See \BMGa\
and references therein} are not used
here. As for the superfield formulation for anomalies in 2D SUGRA,
see refs.\Ref\rGGMTa{\GGMTa}.}

The use of the extended phase space (EPS) of BFV leads to a new
BRST quantization scheme as given in sect.2, nicely avoiding such
a complexity. The EPS of 2D SUGRA is simplified by taking local
Lorentz and Weyl invariant components of spinor fields. The off-shell
nilpotency of BRST charge is ensured from the beginning (at the
classical level) without introducing the redundant variables
appeared in the configuration-space approach[\rHNUa]. The
EPS is yet enough so large that one can easily find out at which stage
and how the gauge dependence appears in specification of the
BRST anomalies. The quantization scheme given here is of course
noncovariant, but the final results of the anomalies turn out to
fall into simple covariant expressions.

Once the EPS is fixed, we apply the general
method\REF\rFIKB{\FIKB}[\rFIKB,\rFIKMa] for analyzing gauge anomaly
as summarized in sect.3. An anomaly can be identified with a
cohomologically nontrivial solution of a set of consistency
conditions\Ref\rWZA{\WZA}
, which is obtained in $\hbar$
expansion by imposing
the super-Jacobi identities on the anomalous commutators for
BRST charge $Q$ and total hamiltonian $H_{\rm T}$. For theories
with reparametrization invariance such as
gravity and string theories the $Q^2$ anomaly is of primary
importance, because it automatically determines the anomaly in
the commutator $[Q\ ,\ H_{\rm T}]$. In the BFV formalism, $Q$ is
directly constructed from
the classically first-class constraints without invoking gauge-fixing
conditions, and therefore the $Q^2$ anomaly can be investigated in a
completely gauge-independent manner. Only the anomalous
$[Q\ ,\ H_{\rm T}]\ $ involves the gauge dependence via the total
hamiltonian, which however can be easily tracked in this formalism.
   It should be stressed that the consistency
 conditions in our formalism[\rFIKB,\rFIKMa] may correspond to
 a hamiltonian version of the descent equations
\REFS\rBZa{\BZa}\REFSCON\rAGGA{\AGGA}
\REFSCON\rStrB{\StrB\hfill\break\ZumA\hfill\break\ZWZ}
[\rBZa,\rStrB], but their physical
 meaning is much more clear than the purely mathematical methods.

In sect.4, we solve algebraically in the full EPS
the consistency condition on the BRST anomalies
to find the most general form of the $Q^2$ anomaly in 2D SUGRA system.
It can be identified with the super-Virasoro anomaly which is
generated as anomalous Schwinger terms in the commutator algebra
of super-Virasoro constraints defined by the BRST transform of
the ghost momenta. The overall factor of the anomaly may be fixed
 by the explicit calculation using, for instance, the normal
ordering prescription. This is the genuine anomaly of the
superstring theory in our formalism, being pregeometric and
determined without any reference to gauge conditions.
In order to identify this $Q^2$ anomaly with
the super-reparametrization or the super-Weyl
anomaly, one must define metric variables
and their superpartners in terms of the BFV basis.  A partial
gauge-fixing is needed to make this geometrization as described
in sect.5.

When the two-dimensional metric is identified,
the relation between the BFV ghost basis and the covariant ghost basis is
 established.  If one expresses the genuine $Q^2$ anomaly in terms of
 the covariant ghosts, this corresponds to a noncovariant anomaly
which spoils
 the super-reparametrization invariance.
  However, it is geometrized so as to respect that symmetry
by adding a suitable
coboundary term given in sect.6. This new expression is the
supersymmetric extension of curvature dependent $Q^2$ anomaly
found in I. At this stage, there remain three bosonic and four
fermionic gauge conditions being left unfixed. They can be used
to make conventionally used gauge-fixings in the configuration space
such as the superconformal, the light-cone and the harmonic gauges
\foot{For a harmonic-gauge formulation
of 2D gravity, see \Ref\rAbeNaka{\AbeNaka} where
some problem in determining the anomaly coeffients is discussed.}.
 However, for a wide class of gauge choices including these,
the gauge dependence does not appear in the $[Q\ ,\ H_{\rm T}]$
anomaly, and its geometrized expression is exactly identified with
the super-Weyl anomaly.
Therefore, it is easy to see some gauge-fixed forms of the BRST anomalies,
 for instance, the $Q^2$ anomaly in the super-orthonormal gauge
[\rOhta,\rIMNUa].
This demonstrates the hierarchical
relationship among the anomalies, which can not clearly be recognized
 in the gauge-fixed approaches to anomalies in superstring or 2D SUGRA
\REFS\rTanA{\TanA}
\REFSCON\rDegrMan{\DegrMan\hfill\break\Degr}
\REFSCON\rGLO{\GLO}
[\rTanA-\rGLO].
We will also give a novel action
 of super-Liouville type, which naturally emerges from the
geometrization of the $Q^2$ anomaly. It converts super-Virasoro
anomaly to super-Weyl anomaly, hence plays the role of
Wess-Zumino-Witten action. We shall summarize our results
in the final section, and give some key formulae in the appendices.

\chapter{\bf BFV formalism of superstring}

 This section describes a new BFV formulation of 2D SUGRA theory
for superstring. We begin by considering the
action\refmark\rGSW\foot{We shall use the two-dimensional conventions;
the flat world-sheet metric is chosen to be $\h_{ab}
={\rm diag}(-1,1)$. The Dirac matrices $\r^a$ are $\r^0=\s_2$ and
$\r^1=i\s_1$ so that $\r^5=\r^0\r^1=\s_3$ with $\s_i$ being Pauli
martices. }
$$\eqalign{
S_{\rm str}=\int d^2\s e\Bigl[&-{1\over2}\Bigl(g^{\a\b}\dl_\a X\dl_\b X
-i\overline\v\r^\a\nabla_\a\v\Bigr) \cr
&-\overline\chi_\a\r^\b\r^\a\v\dl_\b X
-{1\over4}\overline\v\v\chi_\a\r^\b\r^\a\chi_\b\Bigr],\cr
}\eqn\gensayo
$$
where $X^\m$ and $\v^\m$ $(\m=0,1,\cdots,D-1)$ are, respectively,
the bosonic and the fermionic string variables.
This action has reparametrization invariance, local
Lorentz invariace and local supersymmetry.
Moreover, it is invariant under
local Weyl rescalings and local fermionic
transformations.
  Not all of these symmetries can survive in general upon
quantization because of anomalies.
However, it is legitimate to assume that the local Lorentz
invariance remains intact.  This is because
 that the local Lorentz anomaly is known to be always converted
to the Einstein anomaly\Ref\rBZa{\BZa}.
In what follows, therefore, we
shall eliminate the zweibeins from the action
by using the local Lorentz
invariant variables.
We choose the parametrization for the metric variables
$$
\l^\pm=\pm{e_0{}^\pm\over e_1{}^\pm} = {{\sqrt{-g}\pm g_{01}}\over{g_{11}}},
\qquad\x=\ln g_{11},
\eqn\adm
$$
where $e_\a{}^\pm=e_\a{}^0 \pm e_\a{}^1$.
For the fermionic fields, we use rescaled
upper and lower components defined by
$$
\v=\pmatrix{(-e_1{}^-)^{-{1\over2}}\v_-\cr
(e_1{}^+)^{-{1\over2}}\v_+}~,
\qquad
\chi_{\a}=\pmatrix{(e_1{}^+)^{1\over2}\chi_{\a-}\cr
(-e_1{}^-)^{1\over2}\chi_{\a+}}~,
 \eqn\rscsp
$$
and parametrize the gravitino fields as
$$
\n_\pm=(\chi_0\pm\l^\mp\chi_1)_\pm~, \qquad
\L_\pm=4\chi_{1\mp}~.
\eqn\gravab
$$
Note that the local Weyl rescaling
$e_\a{}^a\rightarrow e^\vp e_\a{}^a$
 generates a translation $\x\rightarrow \x+2\vp$
but it leaves $\l^\pm$, $\f_\pm$, $\n_\pm$ and $\L_\pm$
unchanged.
These rescaled spinor components
are local Lorentz and Weyl invariant.\foot{In
ref.\Ref\rKPZ{\KPZ}
similar variables are used to investigate 2D quatum gravity
coupled to Majorana field.} The local fermionic transformation
$\chi_\a\rightarrow \chi_\a+i\r_\a\h$ with $\h$ being an
arbitrary Majorana spinor, on the other hand, induces a change
$\L_\pm\rightarrow \L_\pm-4\h_\pm$ with $\h_\pm$ being the rescaled
local Lorentz and Weyl invariant components of $\h$,
 while $\n_\pm$ remain invariant.
As we shall see in sect.5, the $\L_\pm$ is the superpartner
 of the comformal mode $\x$.
In terms of the variables defined in \adm and \gravab, the action
\gensayo\ can be written as\foot{The world-sheet coordinates
$\s^\a$ ($\a=0,1$) are denoted by $(\t,\s)$, and
take $-\infty < \s < \infty$. It is straightforward to
 make the anlysis on a finite interval of $\s$ so as to impose
the Neveu-Schwarz or Ramond
boudary conditions. We also use $\dot F=\dl_\t F$ and $F'(=\dl F)=
\dl_\s F$ for any variable $F$.}
$$\eqalign{
S_{\rm str}=\int d^2\s\Bigl[{1\over\l^++\l^-}(\dot X&-\l^+X')(\dot X+\l^-X')\cr
&+{i\over2}\v_+(\dot\v_+-\l^+\v_+')
+{i\over2}\v_-(\dot\v_-+\l^-\v_-') \cr
+{2\over\l^++\l^-}\bigl\{&i(\dot X-\l^+ X')\v_-\n_+-i(\dot X+\l^-X')\v_+\n_-
\cr
&+\v_+\v_-\n_+\n_- \bigr\}\Bigr], \cr
}\eqn\shinsayo
$$
Let us denote the canonical momenta for $X$, $\l^\pm$, $\x$, $\n_\pm$
and $\L_\pm$ by $P$, $\p^\l_\pm$, $\p_\x$, $\p_\n^\pm$ and $\p_\L^\pm$,
respectively. Then the canonical theory of this system involves the
following set of primary constraints
$$\eqalign{
&\vp_{\rm A}^{}\equiv \p_{\rm A}^{}\approx0 \qquad
{\rm for~~A}=\l^\pm,\x, \cr
&{\cal J}^z\equiv \p^z\approx0 \qquad {\rm for}~~z=\n_\pm,\L_\pm, \cr
}\eqn\primconst$$
and the secondary constraints, the super-Virasoro constraints
$$\eqalign{
&\vp_\pm={1\over4}(P\pm X')^2\pm{i\over2}\v_\pm\v_\pm'\approx0 \cr
&{\cal J}_\pm=\v_\pm(P\pm X')\approx0. \cr}\eqn\sVconst
$$
They satisfy the classical super-Virasoro algebra
$$\eqalign{
&\{\vp_\pm(\s),\vp_\pm(\s')\}=\pm(\vp(\s)+\vp(\s'))\dl_\s\d(\s-\s'),\cr
&\{{\cal J}_\pm(\s),\vp_\pm(\s')\}
=\pm{3\over2}{\cal J}_\pm(\s)\dl_\s\d(\s-\s')
\pm{\cal J}'_\pm\d(\s-\s'), \cr
&\{{\cal J}_\pm(\s),{\cal J}_\pm(\s')\}=-4i\vp_\pm(\s)\d(\s-\s'),\cr
&{\rm all~other~super\hbox{-}Poisson~brackets~vanish,}
}\eqn\csValg
$$
where $\{~,~\}$ denotes super-Poisson bracket[\rBFVREV].
All the constraints \primconst\ and \sVconst\ are classically
first-class and generate full set of the local symmetries of the
classical action \gensayo.

The BFV argorithm then defines the EPS
 by introducing canonical pairs of the ghost, anti-ghost and
auxiliary fields to each constraint as
$$\eqalign{
\vp_{\rm A}^{}:~~~&({\cal C}^{\rm A},\overline{\cal P}_{\rm A}),\quad
({\cal P}^{\rm A},\overline{\cal C}_{\rm A}),\quad
(N^{\rm A},B_{\rm A})\qquad {\rm for~~A}=\l^\pm,\x,\pm, \cr
{\cal J}^z:~~~&(\g^z,\overline\b_z),\quad
{}~(\b^z,\overline\g_z),\quad
{}~(M^z,A_z) \qquad {\rm for}~~z=\n_\pm,\L_\pm,\pm.
}\eqn\epsval
$$
Here ${\rm A}=\l^\pm,\x,\pm$ and $z=\n_\pm,\L_\pm,\pm$,
respectively, label the bosonic and fermionic first-class
constraints given in \primconst\ and \sVconst. The grassmannian
parities of the ghost variables for the bosonic (fermionic)
constraints are chosen to be odd (even), while those of
auxiliary fields are even (odd).

In the BFV formalism the BRST charge is constructed without fixing
a gauge. It is solely determined by the Poisson algebra among the
constraints and the requirement of nilpotency $\{Q,Q\}=0$. In the case
at hand it takes of the form
$$\eqalign{
Q=\int d\s\Bigl[~&{\cal C}^{\rm A}\vp_{\rm A}^{}+\g^z{\cal J}_z
+{\cal P}^{\rm A}B_{\rm A}+\b^zA_z \cr
&+{\cal C}^+(\overline{\cal P}_+{\cal C}^{+\prime}
+\overline\b_+\g^{+\prime})
+\g^+\Bigl(2i\overline{\cal P}_+\g^+
-{1\over2}\overline\b_+{\cal C}^{+\prime}\Bigr)\cr
&-{\cal C}^-(\overline{\cal P}_-{\cal C}^{-\prime}
+\overline\b_-\g^{-\prime})
+\g^-\Bigl(2i\overline{\cal P}_-\g^-
+{1\over2}\overline\b_-{\cal C}^{-\prime}\Bigr)~\Bigr], \cr
}\eqn\QB
$$
where ${\rm A}$ and $z$ run over all constraint labels.
It generates the BRST transformations of the fundamental variables:
$$\eqalign{
&\d X={1\over 2}\{ ({\cal C}^+-{\cal C}^-)X'+({\cal C}^++{\cal C}^-)P\}
 +\g^+\v_++\g^-\v_-,\cr
&\d P=\Big({1\over2}\{({\cal C}^+-{\cal C}^-)P+({\cal C}^+
+{\cal C}^-)X'\}+\g^+\v_+-\g^-\v_-\Big)',\cr
&\d\v_\pm=\pm{1\over2}{\cal C}^{\pm\prime}\v_\pm\pm{\cal C}^\pm\v_\pm'
+i\g^\pm(P\pm X'),\cr
&\d\l^\pm={\cal C}_\l^\pm, \qquad ~~~\d\x={\cal C}^\x, \qquad~~~~~~
\d{\cal C}_\l^\pm=0, \qquad~ \d{\cal C}^\x=0,  \cr
&\d\n_\pm=-\g^\n_\pm, \qquad~ \d\L_\pm=-\g^\L_\pm, \qquad~
\d\g^\n_\pm=0, \qquad \d\g^\L_\pm=0, \cr
&\d N^{\rm A}={\cal P}^{\rm A}, \qquad~ \d M^z=-\b^z, \qquad~
\d{\cal P}^{\rm A}=0, \qquad \d\b^z=0,  \cr
&\d N^\pm={\cal P}^\pm, \qquad~ \d M^\pm=-\b^\pm, \qquad
\d{\cal P}^\pm=0, \qquad \d\b^\pm=0, \cr
&\d\overline{\cal C}_{\rm A}=-B_{\rm A}, \qquad
\d\overline\g_z=-A_z, \qquad~~ \d B_{\rm A}=0, \qquad \d A_z=0, \cr
&\d\overline{\cal C}_\pm=-B_\pm, \qquad
\d\overline\g^\pm=-A^\pm, \qquad \d B_\pm=0, \qquad \d A^\pm=0, \cr
&\d\overline{\cal P}_{\rm A}=-\vp_{\rm A}, \qquad
\d\overline\b_z=-{\cal J}_z, \cr
&\d{\cal C}^\pm=\pm{\cal C}^\pm{\cal C}^{\pm\prime}-2i(\g^\pm)^2,
\qquad~~~
\d\overline{\cal P}_\pm=-\F_\pm, \cr
&\d\g^\pm=\mp\bigl({1\over2}{\cal C}^{\pm\prime}\g^\pm
-{\cal C}^\pm\g^{\pm\prime}\bigr), \qquad \d\overline\b_\pm=-I_\pm, \cr
&{\rm with}~ {\rm A}=\l^\pm,\x ~~{\rm and}~z=\n_\pm,\L_\pm, \cr
}\eqn\brstr
$$
where $\d F=-\{Q,F\}$ for any $F$. We have introduced the
generalized super-Virasoro operators
$$\eqalign{
\F_\pm&=\vp_\pm\pm2\overline{\cal P}_\pm{\cal C}^{\pm\prime}
\pm\overline{\cal P}_\pm'{\cal C}^\pm
\pm{3\over2}\overline\b_\pm\g^{\pm\prime}
\pm{1\over2}\overline\b_\pm'\g^\pm \cr
I_\pm&={\cal J}_\pm\mp{3\over2}\overline\b_\pm{\cal C}^{\pm\prime}
\mp\overline\b_\pm'{\cal C}^\pm
+4i\overline{\cal P}_\pm\g^\pm \cr
}\eqn\gsVcon
$$
The BRST charge \QB\ and the transformation \brstr\ have a reflection
symmetry in the EPS which we will use below.  All the EPS variables
are devided into those with $(+)$ or $(-)$ indices and the remainings
carring no $\pm$ indices. The symmetry
\foot{In the superconformal gauge, it reduces to a holomolphic
(anti-holomolphic) decomposition of conformal fields and at the
same time a chiral projection on the world-sheet spinors.}
is an invariance under replacements, $\pm \rightarrow
 \mp$, and $\partial_{\s} \rightarrow - \partial_{\s}$.

The canonical structure of the classical theory is now described
in the EPS. In the next section we shall discuss the BRST
anomalies associated with quantization of the theory.

\chapter{\bf BRST anomalies and consistency conditions}

In this section we will review the general description of the
BRST anomalies in the BFV fromalism[\rFIKMa]. When one
applies the BFV formalism to theories with local gauge symmetries,
the basic gauge algebras
reflecting its classical gauge invariance are formulated in terms
of the BRST charge $Q$ by the condition
$$
\{ Q~ ,~ Q\}=0~. \eqn\nil
$$
The gauge symmetries must be consistent with the time development
of the system. This can be expressed as the conservation of $Q$, i.e.,
$$
{\dot Q} \equiv {dQ\over dt} = \{Q~ ,~H_{\rm T}\} = 0, \eqn\qdot
$$
where $H_T$ is the totoal hamiltonian.

Quantization can be acheived formally by replacing super-Poisson
brackets with supercommutators. At the quantum level, however,
these operators must be suitably regularized to become well-defined.
An anomaly arises if \nil\ and \qdot\ can not be maintained
upon quantization. The anomalous terms may be expanded in $\hbar$
as\foot{We will explicitly write $\hbar$ in dealing with $\hbar$
expansion. Otherwise, we use $\hbar=1$.}
$$\eqalign{
[Q~,~ Q] &\equiv i\hbar^2\Omega+{\rm O}(\hbar^3)\cr
[ Q~ ,~ H_{\rm T}] &\equiv {i\over2}\hbar^2\Gamma+{\rm O}(\hbar^3), \cr
}\eqn\anom
$$
It is convenient to distinguish a supercommutator from a naive
one $[~ ,~ ]_0  $ which is defined via classical super-Poisson
bracket $\{~ ,~ \}$, i.e.,
$$
[A~ ,~ B]_0 \equiv i\hbar \{A~ ,~ B\} . \eqn\naivcom
$$
Our basic assumption is that the supercommutation relations between
$Q$ and $H_{\rm T}$ obey the commutation law, the distribution law and
 especially the super-Jacobi identity. They read in the present case
$$
[Q~ ,[Q~ ,~ Q]~ ]=0~ , \eqn\sjacba
$$
$$
 2~ [Q~ ,~ [Q~ ,~ H_{\rm T}]~ ]+[H_{\rm T}~ ,~ [Q~ ,~ Q]~ ]=0~.
\eqn\sjacbb
$$
To the lowest order, ${\hbar}^3$, the outer commutators in
these super-Jacobi identities can be truncated by the naive
commutators, yielding two consistency conditions:
$$
\d \Omega = 0, \eqn\cstome
$$
$$
\d \Gamma = \{H_{\rm T}~ ,~ \Omega\}~  =~ -
{{d\Omega} \over {dt}}~ , \eqn\cstga
$$
where $\d$ is the classical BRST transformation defined
by \brstr.

For any reparametrization invariant theory, the specification of
the BRST anomalies can be considerably simplified. Such a theory
has vanishing canonical hamiltonian, and its total hamiltonian
takes the form, $\displaystyle{H_{\rm T}
=  {1 \over {i \hbar}} [Q~ ,~ \Psi]~}$,
where $\V$ is the gauge fermion\refmark\rFVA\ needed to fix the
gauge degrees of the system. Using \anom\ and the super-Jacobi
identity, we obtain
$$
\Gamma = \{\Omega~ ,~ \Psi\}~ . \eqn\dirgam
$$
One finds that $\Gamma$ can be calculated from $\Omega$ without
solving \cstga. In this sense $\Omega$ is of primary importance for any
 theory being reparametrization invariant at the classical level.
This fact is used below to analyse the BRST anomalies in 2D SUGRA
of superstring theory. It is also noted that the consistency
condition \cstome\ shows up cohomological nature of the anomalies.
If $\Omega$ is a solution of \cstome, then $\bar\W$ defined by
$$
\bar{\Omega} = \Omega + \d \X \eqn\newome
$$
also solves \cstome\ for any $\X$. This is nonvanishing square of
a suitably redefined BRST charge, $[\bar{Q}~,~ \bar{Q}]
= i\hbar^2\bar{\Omega}$, where
$$
   \bar{Q} = Q - {\hbar \over 2}\X~.  \eqn\redefq
$$
The shift of the BRST charge corresponds to a redefinition of the
total hamiltonian as
$$
{\bar H}_{\rm T} = H_{\rm T} - {\hbar \over 2}\{\X,~\Psi \} ~.
\eqn\redefht
$$
It generates an
anomalous commutator $[\bar{Q},~{\bar H}_{\rm T}]$, which defines
$$
\bar{\Gamma} = \{\bar{\Omega}~,~\Psi\} =
\Gamma - {{d \X}\over{dt}} + \d \{\X,~\Psi \}~.
\eqn\newga
$$
One finds from \newome\ and \redefht\ that adding a coboundary
term to a given solution $\Omega$ is related with introducing a
counteraction
$$
 S_{\rm count} = {\hbar \over 2} \int d \tau \{\X,~\Psi \}~,
\eqn\countact
$$
which reflects the difference of underlying regularization schemes
 used to obtain $\Omega$ and $\bar{\Omega}$.

In summary, a BRST anomaly is determined by a cohomology class of
the nontrivial solutions of \cstome,
totally independent of gauge fixings and regularizatrion shemes.
On the other hand, $\Gamma$, the descedant of $\Omega$,
is gauge dependent.
\foot{We regard $\Gamma$ as the descendant of $\Omega$ because
$\Gamma$ is basically determined by $\Omega$ via \dirgam.}
 The gauge dependence
 can be yet easily tracked, and is shown to disappear eventually for
the conventionally used gauge choices.

With these discussions in mind, we shall solve the consistency
condition \cstome\ for 2D SUGRA theory considered in the previous
section.

\chapter{\bf Solution in the extended phase space:
The genuine super-Virasoro anomaly}

In order to solve the consistency condition \cstome\ in the EPS of
2D SUGRA, we will basically follow the method developed in I.
Let us first summarize assignments of ghost number and
canonical dimension for the ghost and auxiliary fields in \epsval.
   The BFV ghosts $ {\cal C}^{\rm A},~ {\cal P}^{\rm A},~\g^z$ and
$\b^z$ carry one unite of the ghost number,
$\rm{gh}({\cal C}^{\rm A}) = \rm{gh}({\cal P}^{\rm A}) =
{\rm gh}(\g^z) = {\rm gh}(\b^z)=1$,
while $~~\rm{gh}(\overline{{\cal P}}_{\rm A}) =
\rm{gh}(\overline{{\cal C}}_ {\rm A}) = {\rm gh}(\overline\g_z)=
{\rm gh}(\overline\b_z)=-1$
for their canonical momenta, $\overline{{\cal P}}_{\rm A}$,
$\overline{{\cal C}}_ {\rm A}$, $\overline\g_z$ and $\overline\b_z$.
Canonical pairs of the auxiliary fields $(N^{\rm A},~B_{\rm A})$ and
$(M^z,~A_z)$ have no ghost number.
We assign $0$ to the canonical dimension of
$X^{\mu}, {\lambda}^{\pm}$ and $\xi$,
and correspondingly $+1$ to $P_{\mu},\pi^{\lambda}_{\pm}$ and
$\pi_{\xi}$.
Their superpartners, $\psi_{\pm},\nu_{\pm},\Lambda_{\pm}$,
and their conjugate momenta have the canonical dimension ${1\over2}$.
  The canonical dimensions of ghosts and anti-ghosts
are fixed only relative to that of ${\cal C}^{\pm}$. Putting $c
\equiv \dim({\cal C}^{\pm})$, we find
$$\eqalign{
&\dim({\cal C}_{\lambda}^{\pm}) = \dim({\cal C}^{\xi})=1+c ,\quad
\dim(\overline{{\cal P}}_{\pm}) = 1-c ,\quad
\dim(\overline{{\cal P}}^{\lambda}_{\pm}) =
\dim(\overline{{\cal P}}_{\xi})=-c, \cr
&\dim(\g^z)=c + {1\over2},\quad\dim(\overline{\b}_z)=-
c + {1\over2}. \cr
}\eqn\ghdim
$$
Note that all the bosonic (anti-)ghosts have the same
canonical dimensions.
\foot{We do not need to know canonical
dimensions of other fields as we will see later.}

  We are now ready to solve the consistency condition
\cstome\ and seek the solution in the form
$$
\Omega = \int d \sigma \omega~ ,
\eqn\omegad
$$
where we assume that $\omega$ is a polynomial of local operators with
${\rm gh}(\omega) = 2$ and $\dim(\omega) = 3+2c$.
According to the general structure of the BFV formalism,
the total phase space
can be divided, with respect to the action of $\d$,
into two sectors;
$$\eqalign{
& {\rm S}_{1}:\ {\rm consisting \ of}\
(X^{\mu}, P_{\mu}, \psi_{\pm})~,~({\cal C}^{\pm} ,
\overline{{\cal P}}_{\pm}) ~~{\rm and}\cr
& {\rm S}_{2}:\ {\rm consisting \ of \ all \ the \ other \ fields}. \cr
}\eqn\sector
$$
It is easy to see that on each sector the $\d$ operation closes:
$$
{\d_{1}}^2 = {\d_{2}}^2 = 0~ ,
\qquad~ \d_{1} \d_{2} + \d_{2} \d_{1}~  =~ 0~ ,
\eqn\decom
$$
where $\d =\d_{1} + \d_{2}$, and $\d_{1} (\d_{2})$ acts only on
${\rm S}_{1}$ (${\rm S}_{2}$) variables.
Since the variables in the ${\rm S}_{2}$-sector form pairs
$(U\ ,\ V)$ with properties $\d_{2} U= \pm V$,
it can be shown that variables belonging to ${\rm S}_2$ can be
removed from $\W$ as coboundary terms\refmark\rFIKB. Therefore,
$\omega$ can be chosen to be independent of the
${\rm S}_2$-variables.

To reduce the number of possibilities for $\w$, we shall use here
the following global symmetries compatible with the BRST
transformation \brstr. The BRST charge \QB\ is invariant under
the space-time Poincar\'e transformation $X^\m\rightarrow \L^\m{}_\n
X^\n+a^\m$. It has also the reflection symmetry, as described in sect.2,
of interchanging the indices of the variables $(+)\leftrightarrow(-)$
and reversing the sign of $\dl_\s$. For simplicity we will denote this
symmetry by $\{+\leftrightarrow-\}$.  We may assume
therefore, without loss of generality,
that $\W$ also respects these symmetries. The translational
invariance forbids $X^{\mu}$ to appear in $\omega$ without
derivatives. It is convenient thus to introduce the variables
$$
Y^{\mu}_{\pm} \equiv (P \pm
X^{\prime})^{\mu} \quad{\rm with}\quad
\d Y^{\mu}_{\pm}  =  \pm \partial_{\sigma}
 ({\cal C}^{\pm} Y_{\pm} + 2\g^{\pm} \psi_{\pm})^\m~.
\eqn\defY
$$
With these  symmetries imposed, there are still great many of
the operators composed of ${\rm S}_1$ variables with ghost number 2
and canonical dimension $2c + 3$. We can, however, reduce futher the
number of operators contributing to $\W$ by noting that the ghost
momenta $\overline{\cal P}_\pm$ and $\overline\b_\pm$ can be removed
from nontrivial solutions to \cstome\ as coboundary terms. We will
give the proof of this lemma in Appendix B. This implies that we
have only to consider the operators composed of ghosts variables
$({\cal C}^\pm,\g^\pm)$ and the string variables $(Y_\pm,\v_\pm)$. These
operators can be classified into the following six groups;
(1) operators constructed only from $({\cal C}^+,\g^+)$, and those
obtained by applying the reflection symmetry $\{+\leftrightarrow-\}$,
(2) operators bilinear in $({\cal C}^+,\g^+)$ and $({\cal C}^-,\g^-)$,
and containing no string variables, (3) operators constructed from
$({\cal C}^+,\g^+)$ and quadratic in $(Y_+, \v_+)$, and those
obtained by $\{+\leftrightarrow-\}$, (4) operators bilinear
in $({\cal C}^+,\g^+)$ and $({\cal C}^-,\g^-)$ and quadratic in
$(Y_+,\f_+)$, and those obtained by $\{+\leftrightarrow-\}$,
(5) operators constructed from $({\cal C}^\pm,\g^\pm)$
and bilinear in $(Y_+, \v_+)$ and $(Y_-,\v_-)$,
(6) operators quartic in the string variables $(Y_\pm,\v_\pm)$.
Since the BRST transformation does not mix operators belonging to
different groups, we can investigate solutions to \cstome\ group by
group. We fisrt assume $\W$ to be the most general linear combinations
of operators belonging to each group and then determine the unkown
coefficients appearing in $\W$ to satisfy the consistency
condition \cstome. In spite of a great many unknown coefficients,
it can be shown that there are no nontrivial solutions in the
cases (2),(3),(4) and (6), and the operators (1) and (5), respectively,
possess only one nontrivial solution $\W_{(1)}$ and $\W_{(5)}$ given by
\def\plram{\{+\leftrightarrow-\}}
$$\eqalign{
\W_{(1)}=\int d\s\Bigl(~&{\cal C}^{+}\dl^3_\s{\cal C}^{+}
-8i\g^{+}\dl^2_\s\g^{+}~\Bigr)+\plram,  \cr
\W_{(5)}=\int d\s\Bigl[~&({\cal C}^{+}\dl_\s{\cal C}^{+}
-{\cal C}^{+}\dl_\s{\cal C}^{-}
-2i(\g^+)^2)Y_+ Y_--4\psi_+\psi_-\g^+\dl_\s\g^- \cr
&+2{\cal C}^{+}(\dl_\s Y_+\psi_++2Y_+\dl_\s\psi_+)\g^-
+2{\cal C}^{+}\psi_+(\dl_\s Y_-\g^++2Y_-\dl_\s\g^+) ~\Bigr] \cr
&+\plram. \cr
}\eqn\nontrivialsols
$$
We thus obtain the general solution to the consistency condition
\cstome
$$
\W=k\W_{(1)}+k'\W_{(5)}. \eqn\genesol
$$
The coefficients $k$ and $k^{\prime}$ can not be determined in
this algebraic approach.  The two nontrivial solutions
\nontrivialsols\ are exactly supersymmetric extension of those in
the bosonic string theory found in I. The first solution $\W_{(1)}$
can be related to the super-Virasoro anomaly as we will see below,
while $\W_{(5)}$ have never been noted before as far as we know.
It depends on string coordinates and is algebraically allowed
anomaly, whose physical implication is not yet unkown. In what
follow we will simply assume
$$
k'=0. \eqn\kprime
$$
The coefficent $k$ can be computed within a specific
regularization scheme. Since $\Omega$ is bilinear in the
ghost fields, it may originate from the anomalous terms in the
gauge algebra for the generalized super-Virasoro constraints \sVconst.
Such anomalous Schwinger terms in the constraint algebra can be
calculated by using the normal ordering prescription in the
Schr\"odinger picture.\foot{See, for
example,\REFS\rCTa{\CTa}[\rCTa,\rManB].} We find
$$\eqalign{
&[\F_\pm(\s),\F_\pm(\s')]=\pm i(\F(\s)+\F(\s'))\dl_\s(\s-\s')
\mp i{D-10\over16\p}\dl_\s^3\d(\s-\s'),\cr
&[I_\pm(\s),\F_\pm(\s')]
=\pm i{3\over2}I_\pm(\s)\dl_\s\d(\s-\s')
\pm iI'_\pm(\s)\d(\s-\s'), \cr
&[I_\pm(\s),I_\pm(\s')]=4\F_\pm(\s)\d(\s-\s')
-{D-10\over2\p}\dl_\s^2\d(\s-\s'),\cr
&{\rm all~other~supercommutators~vanish,}
}\eqn\qsValg
$$
which implies
$$
k=-{D-10\over16\p}.
\eqn\const
$$
Thus one finds that $\Omega$ is essentially unique and
determined without referring to specific gauge choices.

Let us turn to consider $\G$, which is given by a naive commutator
between $\Omega$ and $\Psi$ as in \dirgam. $\G$ can not be calculated
without any assumption on the gauge fermion $\Psi$. First, we
restrict ourselves to the standard form of the gauge fermion[\rBFVREV]
$$
\V=\int d\s[~\overline{\cal C}_{\rm A}\chi^{\rm A}
+\overline\g^z\z_z+\overline{\cal P}_{\rm A}N^{\rm A}
+\overline\b^zM_z~], \eqn\gF
$$
where $\chi$'s and $\z$'s are the gauge-fixing functions.
In order for $N^{\rm A}$ and $M_z$ to be identified with
the multiplier fields, these gauge-fixing functions are assumed
 not to depend on the ghost momenta $\overline{\cal P}_{\rm A}$ and
$\overline\b_z$. They are otherwise arbitrary. These are all the
assumptions we need to compute unambiguously the naive commutator
$\{\Omega,~\Psi\}$ given in \dirgam:
$$
\G=2k~\int d\s(~\dl_\s N^+\dl_\s^2{\cal C}^+
+8i\dl_\s\g^+\dl_\s M^+~)+\plram~.
\eqn\Gsvira
$$
This result is independent of the gauge-fixing
functions ${\chi}$'s and $\z$'s as long as the above assumptions,
which are about the weakest ones imposed on $\Psi$, are satisfied.

Although $\Omega\equiv k\W_{(1)}$ has exactly the same form as the
$Q^{2}$ anomaly of ref.[\rOhta,\rIMNUa], its theoretical content is
much richer.
The $\Omega$, along with its descendant $\Gamma$, exhibits namely
the most general form of anomaly in the extended phase space,
suitably called as the {\em genuine super-Virasoro anomaly}.
It is a pregeometric result because it has been obtained without
any reference to a two dimensional super-metric. The geometrical
meaning of ${\lambda}^{\pm}$ and $\xi$, which is given in \adm,
has disappeared in the EPS, as one can see from their BRST
transformations \brstr; they are no longer related to some metric
variables, since the associated ghosts, ${\cal C}_{\lambda}^{\pm}$
and ${\cal C}^{\xi}$, are by no means the reparametrization ghosts
and the Weyl ghost. The same thing happens in the sector of their
superpartner. So at the present stage, the genuine super-Virasoro
anomaly can not be identified with the super-reparametrization or
super-Weyl anomaly. To distinguish these anomalies from each other
we certainly need a metric.

\chapter{\bf Geometrization}

In the previous section we have obtained the BRST anomalies $\W$
and $\G$ in terms of EPS variables, which do not possess direct
geometrical meaning as they are. To endow them with a geometrical
interpretation, we must specify the gauge conditions to relate the
EPS variables with those of configuration space. This provides us
with the basis needed to geometrize the genuine super-Virasoro
anomaly into the super-Weyl anomaly.

We begin by constructing the BRST gauge-fixed action. Using
the BRST invariant total hamiltonian
$$
H_T= {1\over{i\hbar}} [Q,~\V]~.
\eqn\tH
$$
For the standard form of the gauge fermion \gF, we obtain the BRST
gauge-fixed action \foot{The ghost sector can be simplified by
shifting the gauge fermion \gF\ by $\V\rightarrow \V
+\int d\s[\overline{\cal C}_{\rm A}\dot N^{\rm A}
+\overline\g^z\dot M_z]$. This just cancels the Legendre terms
$\int d^2\s[\overline{\cal C}_{\rm A}\dot{\cal P}^{\rm A}
+\overline\g^z\dot\b_z+B_{\rm A}\dot N^{\rm A}+A^z\dot M_z]$ in
constructing the effective action.}
$$
S=S_{cl}+S_{FP}+S_{gf} \eqn\gfa
$$
where
$$\eqalign{
S_{cl}=\int d^2\s[&P\dot X+{i\over2}\v_+\dot\v_+
+{i\over2}\v_-\dot\v_-
+\p^\l_+\dot\l^++\p^\l_-\dot\l^-+\p_\x\dot\x \cr
&+\p^+_\n\dot\n_++\p^-_\n\dot\n_-+\p^-_\L\dot\L_++\p^-_\L\dot\L_-
-N^{\rm A}\vp_{\rm A}+M_z{\cal J}^z], \cr
S_{FP}=\int d^2\s[&\overline{\cal P}_{\rm A}\dot{\cal C}_{\rm A}
+\overline\b^z\dot\g_z-\overline{\cal C}_{\rm A}\d\chi^{\rm A}
+\overline\g^z\d\z_z-\overline{\cal P}_{\rm A}{\cal P}^{\rm A}
-\overline\b^z\b_z \cr
&-N_+(2\overline{\cal P}_+{\cal C}^{+\prime}
+\overline{\cal P}'_+{\cal C}^++{3\over2}\overline\b_+\g^{+\prime}
+{1\over2}\overline\b'_+\g^+) \cr
&+N_-(2\overline{\cal P}_-{\cal C}^{-\prime}
+\overline{\cal P}'_-{\cal C}^-+{3\over2}\overline\b_-\g^{-\prime}
+{1\over2}\overline\b'_-\g^-) \cr
&-M^+({3\over2}\overline\b_+{\cal C}^{+\prime}+\overline\b'_+{\cal C}^+
-4i\overline{\cal P}_+\g^+) \cr
&+M^-({3\over2}\overline\b_-{\cal C}^{-\prime}+\overline\b'_-{\cal C}^-
+4i\overline{\cal P}_-\g^-)], \cr
S_{gf}=\int d^2\s[&-B_{\rm A}\chi^{\rm A}-A^z\z_z]. \cr
}\eqn\Scl
$$

Because of the presence of the primary constraints \primconst,
 the EPS variables $\l^\pm$ and $\n_\pm$ have lost their original
geometrical meaning as metric variables and gravitino fields,
 as discussed in the previous section.
Instead the auxiliary fields $N^\pm$ and $M^\pm$ play that
role as is seen from \Scl.
For the geometrization, these two sets of variables must be identified
 by imposing the gauge conditions\refmark\rFIKMa
$$
\chi_\l^\pm=\l^\pm-N^\pm, \qquad \z^\n_\pm=\n_\pm\mp iM^\mp. \eqn\gcond
$$
There still remain the gauge conditions $\chi^\x$, $\chi^\pm$,
$\z^\L_\pm$ and $\z^\pm$ left unfixed.  These are
for the gauge degrees of freedom, which we would like to
identify with two reparametrization symmetries, one Weyl symmetry
and their supersymmetric counterparts.
  In order to construct the covariant ghost fields from the BFV
 ghosts by using some equations of motion, one needs in general to
fix these gauge symmetries.  However, It suffices for us
to assume
 that the
gauge-fixing functions $\chi^\x$, $\chi^\pm$,
$\z^\L_\pm$ and $ \z^\pm$ and their BRST transforms do not depend on
$P$, $\p_{\rm A}$, $\p^z$, $\overline{\cal P}_{\rm A}$ and
$\overline\b^z$.
This is  the only assumption we have to make for the general analysis.
The gauge conditions can be arbitrary otherwise.

{}From the action \gfa\ we obtain
$$\eqalign{
& \dot X-{1\over2}\{(N^++N^-)P+(N^+-N^-)X'\}+M^+\v_++M^-\v_-=0, \cr
& \dot\v_\pm\mp N^\pm\v'_\pm\mp{1\over2}N^\pm{}'\v_\pm
+iM^\pm(P\pm X')=0, \cr
& \dot\l^\pm=N^\pm_\l, \qquad \dot\x=N^\x, \qquad
\dot\n_\pm=M^\n_\pm, \qquad \dot\L_\pm=M^\L_\pm, \cr
& \dot{\cal C}_\l^\pm={\cal P}^\pm_\l,\qquad
\dot{\cal C}^\x={\cal P}^\x, \qquad
\dot \g^\n_\pm=\b^\n_\pm, \qquad \dot\g^\L_\pm=\b^\L_\pm, \cr
& {\cal P}^\pm=\dot{\cal C}^\pm\pm{\cal C}^\pm N^\pm{}'
\mp{\cal C}^\pm{}'N^\pm-4i\g^\pm M^\pm, \cr
& \b^\pm=\dot\g^\pm\pm{1\over2}\g^\pm N^\pm{}'\mp\g^\pm{}'N^\pm
\mp{\cal C}^\pm M^\pm{}'\pm{1\over2}{\cal C}^\pm{}'M^\pm. \cr
}\eqn\eom
$$
To identify the covariant ghost variables we consider the BRST
transformation of $X$ given in \brstr. Using the equation of motion
for $P$ in \eom, we find on the EPS basis
$$\eqalign{
\d X={{\cal C}^++{\cal C}^-\over N^++N^-}\dot X
&+{N^-{\cal C}^+-N^+{\cal C}^-\over N^++N^-}X' \cr
+\Bigl(\g^+&+{{\cal C}^++{\cal C}^-\over N^++N^-}M^+\Bigr)\v_+
+\Bigl(\g^-+{{\cal C}^++{\cal C}^-\over N^++N^-}M^-\Bigr)\v_-.\cr
}\eqn\dX
$$
On the other hand the BRST transformation in terms of covariant ghosts
is given by
$$
\d X=C^\a\dl_\a X+\overline\w\v = C^\a\dl_\a X-i(\w_-\v_+-\w_+\v_-),
\eqn\cdX
$$
where $C^\a$ ($\a=0,1$) and $\w_\pm$ are the ghost varaibles for the
reparametrization and local supersymmetry. The covariant bosonic ghost
is defined by
$$
\w=\pmatrix{(e_1{}^+)^{1\over2}\w_-\cr(-e_1{}^-)^{1\over2}\w_+}.
\eqn\omegagh
$$
By comparing \dX\ with \cdX\ we obtain
$$
C^0={{\cal C}^++{\cal C}^-\over N^++N^-}, \qquad
C^1={N^-{\cal C}^+-N^+{\cal C}^-\over N^++N^-}, \qquad
\w_\pm=\mp i\Bigl(\g^\mp
+{{\cal C}^++{\cal C}^-\over N^++N^-}M^\mp\Bigr).
\eqn\csgh
$$

The Weyl ghost $C_\ssW$ can be found by considering the BRST
transformation of $\x=\ln g_{11}$. In the configuration space it is
given by
$$
\d\x=C_\ssW+C^\a\dl_\a\x+2C^{1\prime}+C^0{}'(\l^+-\l^-)
-i(\w_-\L_+-\w_+\L_-). \eqn\dxi
$$
Since $\d\x={\cal C}^\x$ on the EPS basis and we are working
with the gauge conditions \gcond, we can solve \dxi\ for $C_\ssW$
to obtain
$$
C_\ssW={\cal C}^\x-V^+_{\cal C}+V^-_{\cal C}, \eqn\cwgh
$$
where $V_{\cal C}^\pm$ and their companions $V_N^\pm$ are defined by
$$\eqalign{
&V_N^\pm={\, 1\,\over2}G_\pm N^\pm\pm\L_\pm M^\pm+N^{\pm\prime}, \cr
&V_{\cal C}^\pm={1\over2}G_\pm{\cal C}^\pm\pm\L_\pm\g^\pm
+{\cal C}^{\pm\prime} \cr
}\eqn\WtoU
$$
with
$$
G_\pm={2\over N^++N^-}[\pm N^\x+N^\mp\x'\mp(N^+-N^-)'
\mp(\L_+M^++\L_-M^-)].\eqn\gpm
$$
We have replaced the time derivative $\dot\x$ with $N^\x$ by way of
\eom.

The covariant ghost for the fermionic symmetry can also be obtained
from the BRST transformation of $\L_\pm$ in the configuration space
$$\eqalign{
\d\L_\pm=-4\h_{\ssW\pm}&+C^\a\dl_\a\L_\pm-4C^0{}'\n_\pm
+{1\over2}(C^1{}'\pm\l^\pm C^0{}')\L_\pm \cr
&\pm{4g_{11}\over\sqrt{-g}}\w_\mp\{\dot\x\pm\l^\mp\x'-(\l^+-\l^-)'
-i(\n_+\L_--\n_-\L_+)\}+4\w_\mp', \cr
}\eqn\cfsgh
$$
where $\h_{\ssW\pm}$ stand for the ghost for the fermionic
symmetry. In
spinor notation the covariant bosonic ghost for the fermionic symmetry
is given by
$$
\h_\ssW=\pmatrix{(-e_1{}^-)^{-{1\over2}}\h_{\ssW-}\cr
(e_1{}^+)^{-{1\over2}}\h_{\ssW+}}.
\eqn\fcfsgh
$$
Comparing \cfsgh\ with the relation
$\d\L_\pm=-\g_\pm^\L$ in the EPS, we obtain
$$
\h_{\ssW\pm}={1\over4}\Bigl(W_{\cal C}^\pm
+{{\cal C}^++{\cal C}^-\over N^++N^-}W_N^\pm\Bigr), \eqn\apm
$$
where $W_{\cal C}^\pm$ and $W_N^\pm$ are defined by
$$\eqalign{
&W_N^\pm=M^\L_\pm\mp\L'_\pm N^\pm\mp{1\over2}\L_\pm N^\pm{}'
\pm i(G_\pm M^\pm+4M^\pm{}'),\cr
&W_{\cal C}^\pm=\g^\L_\pm\mp\L'_\pm{\cal C}^\pm
\mp{1\over2}\L_\pm{\cal C}^\pm{}'\pm i(G_\pm\g^\pm+4\g^\pm{}'). \cr
}\eqn\wpm
$$
Here the time derivatives $\dot\L_\pm$ have been replaced by $M^\L_\pm$.

The equations \csgh, \cwgh\ and \apm\ completely
fix the ghost relations between the BFV basis and the covariant one.
It is straighforward to obtain the BRST transformation for the
configuration space variables. We give a complete list for the
BRST transformations in covariant form in Appendix A. However, in
the next section we shall use the original form \brstr\ written
in terms of the BFV basis. This is because that the transformation
rule in the EPS, which clearly satisfies the nilpotency condition,
is particularly simple compared with those for the configuration
space variables. This makes the general analysis much easier.
The ghost relations found above are used only after performing the
BRST transformation to obtain the manifestly covariant expression
of the super-Weyl anomaly.

\chapter{\bf Derivation of the super-Weyl anomaly}

This section describes the central issue of the present paper,
derivation of the super-Weyl anomaly from the genuine
super-Virasoro anomaly found in section 4.

In the previous section we have shown that the original EPS variables
are related to covariant geometrical ones after the partial gauge
fixing specified by \gcond.
When expressed in terms of these covariant ghosts via \csgh,
the genuine Virasoro anomaly $\Omega$ and $\Gamma$ might be interpreted as
 an anomaly associated with super-reparametrizations.
One expects, however, that this anomaly
may be shifted into the one which respects the super-reparametrization
 symmetry and is endowed with a geometrical interpretation.
As we shall show in this section, this can, indeed, be acheived and
the geometrized anomaly turns out to be the super-Weyl anomaly.

Let us denote the geometrized $Q^2$ anomaly and its descendant by
${\Omega}_{g}$ and $~{\Gamma}_{g}$, respectively. They must be
expressed in terms of covariant variables and have well-defined
transformation properties under the covariant BRST transformations
given in Apendix A. As we have argued in sect.4, these two different
expressions for $Q^2$ anomaly, $~\Omega$ given in \genesol\ with $k'=0$ and
${\Omega}_{g}$, should necessarily belong to the same cohomology
class defined by the BRST transformation \brstr\ in the extended
phase space. Therefore, the difference between $\Omega$ and
${\Omega}_{g}$ is proportional to a coboundary term. Such
coboundary term, however, can be traced back to the redefinition
of the BRST charge as discussed in sect.3. Let $Q_g$ stand for
the BRST charge producing the geometrized anomalies $\W_g$ and $\G_g$. Then
it is related with the original BRST charge $Q$ by
$$
Q_g=Q-{\hbar\over2}\X. \eqn\gbrstc
$$
The construction of $\X$ proceeds with some guess work. It must be
a suitable supersymmetric generarization of the corresponding
quantities obtained in I. Such a $\X$ exists and is given by
$$\eqalign{
\X &=k\int d\s\biggl[\, {\, 1\,\over 2} {\cal C}^\xi (~ G_+-G_-
)~ +U_+^{\cal C} + U_-^{\cal C} \biggr] ,\cr
}\eqn\cobaundary
$$
where $U_{\pm}^{\cal C}$ and their companions $U_\pm^N$ are defined by
$$\eqalign{
U_{\pm}^{\cal C}=
&-{\cal C}^\pm\biggl\{ ({\,1 \,\over 4}G_\pm^2-G_\pm^\prime )
                      \pm {i\over\, 2\,}\L_\pm\L_\pm^\prime\biggr\}
+{i\over\, 2\,}\g_\pm^\L\L_\pm\mp\g^\pm (G_\pm \L_\pm -4\L_\pm^\prime),\cr
U_{\pm}^N=
&-N^\pm\biggl\{ ({\,1 \,\over 4}G_\pm^2-G_\pm^\prime )
                      \pm {i\over\, 2\,}\L_\pm\L_\pm^\prime\biggr\}
-{i\over\, 2\,}M_\pm^\L \L_\pm\pm M^\pm (G_\pm \L_\pm -4\L_\pm^\prime).\cr
}\eqn\cobaundarybf
$$
In deriving \cobaundary\ we have used the assumption on the
gauge-fixing functions mentioned in sect.5. It is now straightforward
to obtain $\W_g$ and $\G_g$. On the BFV basis, the $~{\Omega}_{g}$ is
given by
$$\eqalign{
\W_g=k\int d\s\Biggl[
&{{{\cal C}^++{\cal C}^-}\over {N^+ +N^-}}\bigl\{ {\, 1\,\over 2}
\dl_\t( G_+-G_-) -\dl_\s( V_N^++V_N^-)\bigr\}
({\cal C}^\x -V^+_{\cal C}+V^-_{\cal C})\cr
+&\biggl\{{2\over {N^++N^-}}\dl_\t({\cal C}^\x -V^+_{\cal C}
+V^-_{\cal C})
-{{N^+-N^-}\over {N^++N^-}}\dl_\s({\cal C}^\x -V^+_{\cal C}
+V^-_{\cal C})\cr
&-{2\over {N^++N^-}}(\g^+ W_N^+ +\g^- W^-_N+M^+ W_{\cal C}^++M^-
W_{\cal C}^-)
\biggr\} \cr
&\times ({\cal C}^\x -V^+_{\cal C}+V^-_{\cal C})
-{\, i\,\over 2}\left\{( W_{\cal C}^+)^2 +(W_{\cal C}^-)^2 \right\}
\Biggr],\cr
}\eqn\test
$$
where $V_N^\pm$ are given in \WtoU.
To show that $\W_g$ indeed be equivalent to $\W$, we briefly describe
how to get $\W_g$ in Appendix C. Using the relations
$$\eqalign{
eR+4i\e^{\alpha \beta}\partial _\alpha
( \overline \chi _\beta \rho _5\rho ^\gamma \chi _\gamma )
  &={\, 1\,\over 2}\dl_\t( G_+-G_-) -\dl_\s( V_N^++V_N^-) \cr
2\e ^{\a\b}\overline\w\r^0\r^5\nabla_\a\chi_\b
&={\, 1 \, \over {N^+ +N^-}}(\w_-W_N^+-\w_+W_N^-),\cr
e\overline{\h}_\ssW\r^\a\r^0\chi_\a
  &={2\over {\, N^++N^-\,}}(M^+\h_{\ssW+}^{}+M^-\h_{\ssW-}^{} ),\cr
e\overline\h_\ssW\r^0\h_\ssW^{}
  &=\h_{\ssW+}^2+\h_{\ssW-}^2,\cr
16\e^{\alpha \beta }\overline \h_\ssW \rho _5
\nabla _\alpha \chi _\beta
  &=i( W_N^+W_{\cal C}^++W_N^-W^-_{\cal C}), \cr
}\eqn\util
$$
and, \csgh\ and \apm\ , one obtains the covariant expression
$$\eqalign{
\W_g=k\int d\s\biggl[
&\{eR+4i\e^{\a\b}\dl_\a(\overline {\chi }_\b\r_5\r^\g\chi_\g)\}
C^0 C_\ssW +eg^{0\a} C_\ssW\dl_\a C_\ssW\cr
&+( 4i\e ^{\a\b}\overline\w\r^0\r^5\nabla_\a\chi_\b
-4e\overline{\h}_\ssW\r^\a\r^0\chi_\a) C_\ssW\cr
&-8ie\overline{\h}_\ssW\r^0\h_\ssW^{}
-16C^0\e^{\a\b}\overline{\h}_\ssW\r_5\nabla_\a\chi_\b\biggr].\cr
}\eqn\curent
$$
where $R$ is the scalar curvature for the metric $g_{\a\b}$
and $e\equiv\sqrt{-g}$.
Once $\W_g$ is given on the BFV basis, it is straightfoward
to calculate $~{\Gamma}_{g} = \{\W_g,~\Psi \}$.  In this
 naive commutator, the ghost fields are simply replaced by
 the relevant multiplier field due to our assumption.
We thus obtain by using the identity, $V_N^+-V_N^-=N^\x$,
$$\eqalign{
\G _g &=-k\int d\s\biggl[ \bigl\{
{\, 1\,\over 2}\dl_\t( G_+-G_-) -\dl_\s( V_N^++V_N^-)\bigr\}
({\cal C}^\x -V^+_{\cal C}+V^-_{\cal C})\cr
&\qquad\qquad\qquad
+i( W_N^+W_{\cal C}^++W_N^-W^-_{\cal C})\biggr] \cr
&=-k\int d\s\biggl[ \{
eR+4i\e^{\alpha \beta}\partial _\alpha
( \overline \chi _\beta \rho _5\rho ^\gamma \chi _\gamma )\}
C_\ssW+16\e^{\alpha \beta }\overline \h_\ssW\rho _5
\nabla _\alpha \chi _\beta\biggr] ,\cr
}\eqn\gammabfvbasis
$$
This is nothing but the super-Weyl anomaly. One may also confirm
that this expression is obtained by adding to \Gsvira\ the
contribution from the coboundary term $\X$ as
$$
{\Gamma}_{g} = \Gamma - ~ \dot{\X} + ~ \d ( \{ \X  , \Psi \} )~.
\eqn\altgam
$$
The results \curent\ and \gammabfvbasis\ are just the supersymmetric
generalization of the geometrized BRST anomalies obtained in I.

Our covariant expressions \curent\ and \gammabfvbasis\ are invariant
under supersymmetry transformation while the Weyl invariance and the
fermionic symmetry are necessarily broken upon quantization unless
$D=10$. We stress that derivation of the super-Weyl anomaly
given here is by construction nonperturbative and practically
gauge independent. At first sight the gauge independence of $\G_g$
might be considered to be odd since it is directly related with the
gauge fermion as in \dirgam. As we have noted in sect.4, this
peculiar property can be understood if one notice that the gauge
conditions $\chi$'s and $\z$'s do not contribute to the rhs of
\dirgam\ so far as the assumptions on the gauge conditions
mentioned in sect.5 are satisfied.

To make it clear the implication of geometrization,
it is interesting to note that the genuine super-Virasoro anomaly
\genesol\ implies the breakdown of the reparametrization invariance
and the local supersymmery of the classical action \gensayo\ with the
superghost sector included, while the Weyl rescaling and the fermionic
symmetry remain intact.
As we have shown in this section, it can be
converted to the super-Weyl anomaly by a suitable redefinition of the
BRST charge given by \gbrstc. Then the geometrized BRST charge $Q_g$
nesessarily respects the reparametrization invariance and the local
supersymmetry.
This can most easily be seen from the counteraction
associated with the transition from $Q$ to $Q_g$. Let us denote the
counteraction by $S_g$, then it is given by
$$\eqalign{
S_{g}  &= {1\over2}\int d \tau \{\Xi, ~\Psi\} \cr
       &= {k\over2}\int d^2 \sigma
\Bigl[{1\over2}N^{\x} (G_{+} - G_{-})+U_+^N + U_-^N \Bigr] .\cr
}\eqn\sg
$$
In terms of configuration space varibles, \sg\ can be written as
$$\eqalign{
S_g=-{k\over2}\int d^2\s \Bigl[&e\Bigl\{-{1\over2}\Bigl(
g^{\a\b}\dl_\a\x\dl_\b\x-i\overline\L\r^\a\nabla_\a\L\Bigr)
-\overline\chi_\a\r^\b\r^\a\L\dl_\b\x
-{1\over4}\overline\L\L\chi_\a\r^\b\r^\a\chi_\b\Bigr\} \cr
&+eR\x+4i\e^{\a\b}\overline\chi_\a\r_5\r^\g\chi_\g\dl_\b\x
+4\e^{\a\b}\overline\chi_\a\r_5\nabla_\b\L
-{2g_{11}\over\sqrt{-g}}\Bigl\{\Bigl({g_{01}\over g_{11}}\Bigr)'\Bigr\}^2
\Bigr],\cr
}\eqn\slgensayo
$$
where $\L$ is defined by
$$
\L=\pmatrix{(-e_1{}^-)^{-{1\over2}}\L_-\cr
(e_1{}^+)^{-{1\over2}}\L_+\cr}.
\eqn\largelabmda
$$
Since both $\x$ and $\L$ do not possess simple transformation
properties like scalars and spinors under reparametrization
and local supersymmetry, $S_g$ is not invariant under
these symmetries. Instead, it exactly
cancels the super-Virasoro anomaly of the action $S_{cl}+S_{FP}$
given in \Scl \Ref\rFSa{T. Fujiwara and T. Suzuki, in preparation}.
Under the Weyl rescaling $\displaystyle{\d_\vp e_\a{}^a
={\vp\over2}e_\a{}^a}$
and $\displaystyle{\d_\vp\chi_\a={\vp\over4}\chi_\a}$, and the
fermiomic symmetry $\d_\h e_\a{}^a=0$ and $\d_\h\chi_\a=i\r_\a\h$ with
$\h$ being an arbitrary Majorana field, $\x$ and $\L$ behave as
super-Liouville mode. Furthermore, $S_g$ correctly reproduces the
super-Weyl anomaly relations
$$\eqalign{
\d_\vp S_g&=-{k\over2}\int d^2\s[eR
+4i\e^{\a\b}\dl_\a(\overline\chi_\b\r_5\r^\g\chi_\g)]\vp, \cr
\d_\h S_g&=-8k\int d^2\s\e^{\a\b}\overline\h\r_5\nabla_\a\chi_\b \cr
}\eqn\superWeyl
$$
In this sense $S_g$ is nothing but the super-Liouville action with
$\x$ and $\L$ as the super-Liouville fields. It is a local functional
of 2D supergravity fields in sharp contrast to the nonlocal
super-Liouville action\REFS\rGXa{\GXa}[\rBMGa,\rGGMTa,\rGXa]. This
is olny possible by sacrificing the reparametrization invariance
and the local supersymmetry. Since the super-Virasoro anomaly is
converted to the super-Weyl anomaly by introducing $S_g$ as a
counteraction, it can be regarded as a Wess-Zumino-Witten
term[\rWZA,\rBZa].

Another point to be noted in connection with \slgensayo\ is that
the super-Liouville mode of the 2D supergravity becomes propagating
if one include \slgensayo\ to \gensayo\ as was discussed in
ref.[\rPolA]. Some of the classical constraints generating the
super-Weyl symmetry will be lost from \primconst, and the
contributions from the 2D supergravity sector should be included in
\sVconst. The issues related to this has been discussed
in \Ref\rFIKMTa{\FIKMTa, \hfill\break\FIKTb} for the bosonic string.
The extension to fermionic string will be discussed in
\Ref\rFIKTa{T. Fujiwara, Y. Igarashi, R. Kuriki and T. Tabei, in
preparation}.

With these observations in mind it is instructive to discuss some
gauge-fixed versions of the BRST anomalies. We examine the
superconformal gauge[\rDZa,\rKOa,\rIMNUa] and the
supersymmetric extension of the light-cone gauge\REFS\rSie{\SieA\hfill
\break\Fuk}
\REFSCON\rPolB{\PolB}
\REFSCON\rPZa{\PZa\hfill\break\AAZa}[\rSie,\rPolB,\rGXa,\rPZa].
\foot{The light-cone
coordinates are defined by $\s^\pm=\t\pm\s$. For derivatives we employ
the convention $\dl_\pm=\dl_\t\pm\dl_\s$.}

\noindent
(i) Superconformal gauge

In addition to \gcond\ this gauge is realized by the following
choice of the gauge conditions which are the simple extension
of the bosonic string case[\rFIKMa]
$$\eqalign{
&\chi^\pm\equiv N^{\pm}-1, \qquad \chi^\xi\equiv\xi-{\hat \xi},\cr
&\z^\pm\equiv iM^\pm,\qquad~~~~
\z^\L_\pm\equiv{\Lambda}_{\pm}-{\hat{\Lambda}}_{\pm}, \cr
}\eqn\conformal
$$
where ${\hat \xi}$ and ${\hat{\Lambda}}_{\pm}$ are fixed functions.
This corresponds to the choice, $g_{\a \b} = \eta_{\a \b}
\exp({\hat \xi})$ and $\displaystyle{\chi_{\a}
= {i\over4}\rho_{\a} {\hat{\L}}}$. The Weyl
ghost\Ref\rFujG{\FujG} and its superpartner can be then eliminated via
the equations of motion
$$\eqalign{
  C_\ssW =& - {1\over2}[(C^{+} {\dl}_{+} +C^{-} {\dl}_{-}){\hat{\xi}}
 +({\dl}_{+} C^{+} + {\dl}_{-} C^{-})
+i({\L}_{+} {\w}_{-} -{\L}_{-} {\w}_{+}), \cr
 \eta_{\ssW\pm} =& {1\over8} (C^{+} {\dl}_{+} +
C^{-} {\dl}_{-}){\hat{\L}}_{\pm} \pm {1\over16}{\dl}_{\pm} C^{\pm}
 {\hat{\L}}_{\pm}\pm\w_{\mp}{\dl}_{\pm}{\hat \xi}\pm{1\over2}
{\dl}_{\pm}\w_{\mp}, \cr
}\eqn\Weylghost
$$
where the reparametrization ghost, $C^{\pm}=C^{0}+C^{1}$, and its
partner, $\w_{\pm}$, satisfy ${\dl}_{\pm} C^{\mp}=0$ and ${\dl}_{\pm}
\w_{\pm}=0$. It is easy to compute gauge-fixed forms of $\W_g$
and $\G_g$, which depend on $\hat\x$ and $\hat\L_\pm$ in \conformal.
 Especially in the super-orthonormal gauge with ${\hat \xi}
={\hat{\L}}_{\pm}=0$ we obtain
$$\eqalign{
 \W_{g}=& k \int d\s [C^{+} \dl^3 C^{+} + 8i \w_{+} \dl^2 \w_{+}]
 + \plram, \cr
  \G_{g}=& 0. \cr
}\eqn\WGortho
$$
This is the supersymmetric extension[\rOhta,\rIMNUa] of the result
of Kato-Ogawa[\rKOa], and corresponds to the BRST gauge-fixed
version of the genuine super-Virasoro anomaly. The super-Weyl anomaly
can not directly be seen in this special gauge. One should not conclude
from this result the absence of super-Weyl anomaly. As noted
in I, it is not legitimate to fix all the
classical gauge degrees if some of the local symmetries become
anomalous. In the present case we can not gauge-fix the
superconformal mode of 2D supergravity due
to the super-Weyl anomaly. It can be shown that one can recover the
super-Weyl anomaly if the contributions of the superconformal mode
to $\G_g$ are properly taken into account.

\noindent
(ii) Light-cone gauge

The supersymmetric extension of the light-cone gauge fixing for the
bosonic string can be defined by
$$
(e_\a{}^a)=\pmatrix{
e_+{}^+ & e_+{}^- \cr
e_-{}^+ & e_-{}^- \cr}
=\pmatrix{
1 & -g_{++} \cr
0 & 1 \cr}, \qquad
\chi_-=\pmatrix{\chi_{--} \cr \chi_{-+} \cr}=0. \eqn\lcgcond
$$
In terms of BFV variables this is equivalent to the following set
of gauge conditions
$$\eqalign{
\chi^+&\equiv N^+-1, \qquad \chi^-\equiv e^\x(N^-+1)-2, \cr
\z^+&\equiv iM^+, \qquad~~~~ \z^-\equiv iM^--{1\over4}(N^-+1)\L_- \cr
}\eqn\lcgbfv
$$
togehter with \gcond. As in the conformal gauge fixing we should
specify $\chi^\x$ and $\z^\L_\pm$ to copmpletely fix the classical
gauge symmetries corresponding to the Weyl rescaling and fermionic
symmetry. Since they are broken by anomalies, we will leave $\chi^\x$
and $\z^\L_\pm$ unspecified.

The equations of motion for spinor components become simple in
terms of rescaled variables: $(\pm e_1{}^\pm)^{1\over2}\w_\mp
\rightarrow \w_\mp$, $(\mp e_1{}^\mp)^{-{1\over2}}\h_{\ssW\mp}
\rightarrow \h_{\ssW\mp}$ and $(\pm e_1{}^\pm)^{1\over2}\chi_\mp
\rightarrow \chi_\mp$. Then the equations of motion for ghosts are
given by
$$\eqalign{
&\dl_-C^+=0, \qquad C_\ssW=-{1\over2}(\dl_+C^++\dl_-C^-)
+4i\w_-\chi_{+-}, \cr
&\dl_-\w_-=0, \qquad
\h_{\ssW-}=-{1\over2}\dl_-\w+, \cr
}\eqn\gheom
$$
while the equations for $C^-$ and $\h_{\ssW+}$ can not be derived
from the gauge conditions \lcgbfv. Using \gheom, we can easily find
the BRST anomalies in the light-cone gauge. In particular, $\G_g$ is
given by
$$\eqalign{
\G_g=&{k\over2}\int d\s(\dl_+C^++\dl_-C^-)\dl_-^2g_{++}\cr
&+8ik\int d\s(\dl_-\w_+\dl_-\chi_{++}
+{1\over2}\w_-\chi_{+-}\dl_-^2g_{++})
-16ik\int d\s\h_{\ssW+}\dl_-\chi_{+-} \cr
=&-{k\over2}\int d\s C^-\dl_-^3g_{++}
-8ik\int d\s\w_+\dl_-^2\chi_{++} \cr
&-16ik\int d\s\Bigl(\h_{\ssW+}+{1\over4}\w_-\dl_-g_{++}\Bigr)
\dl_-\chi_{+-}+\cdots~. \cr
}\eqn\ggilcg
$$
In the second equality we have omitted the total time derivatives.

\chapter{\bf Summary}
Applying the generalized hamiltonian
formalism of Batalin, Fradkin
and Vilkovisky,
we have quantized superstring
theory of Ramond-Neveu-Schwarz to perform
an exhaustive algebraic analysis on anomalies in the EPS.
  To make the analysis most general,
we have presented a new
canonical formulation of 2D SUGRA theory.
  On the basis of this BRST scheme, the genuine super-Virasoro
anomaly expressed
by $\Omega$ is identified
with the essentially unique
solution of the consistency condition,
 $\d \Omega = 0$, without invoking any particular gauge for the
metrics and gravitinos on the world-sheet.
 The absolute normalization for the $\Omega$ can be fixed
by using the canonical normal ordering prescription.
Our analysis shows up
the primary importance of the
genuine super-Virasoro anomaly;
 it is totally gauge independent, and
 the super-Weyl anomaly in the configuration
space is obtained from it as a descendant.
We have derived the most general form of
 the super-Weyl anomaly
by making a partial gauge-fixing and explicitly
finding a local counterterm needed for the covariantization.
The conditions under
which this expression can be independent
of the remaining gauge choices are clarified.
Our results are obtained in a nonperturbative way
 without assuming the
weak gravitational field as in ref.[\rAGWa, \rTanA], and valid in any
 space-time dimensions.
  It is straightforward
to give gauge-fixed forms of these BRST anomalies.
We have examined superconformal gauge and supersymmetric light-cone
gaug as particularly interesting cases.

The above results are summarized as a
 hierarchial relationship among the anomalies
 in fermionic string theory.
In the unconstrained EPS,
the genuine super-Virasoro anomaly is
sitting on the top of the hierarchy of anomalies. In its
subspace, where the two dimensional metric variables and
 their superpartner can be identified,
this pregeometrical anomaly obtains its geometrical meaning
and appears as the super-Weyl anomaly.
 The $Q^2$ anomaly being supersymmetric extension[\rOhta,\rIMNUa] of
the one considered
by Kato and Ogawa[\rKOa] for bosonic string
is obtained as a complete gauge-fixed form of the anomaly in the
super-orthonormal gauge.
  The relationship clarified here between the super-Virasoro and
the super-Weyl anomaly
should be compared with
that discussed in the algebraic
approaches [\rManB, \rDegrMan, \rGLO] using
 descent equations for cocycles[\rBZa-\rStrB].
There, the super-Virasoro anomaly has been
calculated as a 2-cocyle
from 1-cocyle, the super-Weyl anomaly.
  The hierarchial relationships discussed
here can not be revealed in the
 gauge-fixed analyses.

Our result of the gunuine super-Virasoro anomaly is the starting
point for quantizing subcritical fermionic string theory or 2D SUGRA
as an anomalous gauge theory[\rFIKMTa,\rFIKTa]. A systematic
construction
of the super-Liouvill action for the theory will be discussed in a
forthcomming paper[\rFSa].

\vfill\eject
\APPENDIX{A}{A}
\centerline{\bf BRST transformations in the configuration space}
In order to establish the ghost relations between
the BFV and the covariant
 basis, we give the BRST transformation in the configuration space:
$$\eqalign{
& \d X=C^\a\dl_\a X-i(\w_-\psi_+-\w_+\psi_-), \cr
& \d\psi_\pm=C^\a\dl_\a\psi_\pm
+{1\over2}(C^{1\prime}\pm\l^\pm C^{0\prime})\psi_\pm \cr
&\phantom{\d\psi_\pm=C^\a\dl_\a}
\pm{2\over\l^++\l^-}\w_\mp\{\dot X\pm\l^\mp X'
-i(\n_-\psi_+-\n_+\psi_-)\}, \cr
& \d\l^\pm=C^\a\dl_\a\l^\pm\pm(\dot C^1\pm\l^\pm\dot C^0)
-\l^\pm(C^{1\prime}\pm\l^\pm C^{0\prime})-4i\w_\mp\n_\mp, \cr
& \d\x=C_\ssW+C^\a\dl_\a\x+2C^{1\prime}+C^0(\l^+-\l^-)
-i(\w_-\L_+-\w_+\L_-), \cr
& \d\n_\pm=C^\a\dl_\a\n_\pm+(\dot C^0\pm\l^\mp C^{0\prime})\n_\pm
-{1\over2}(C^{1\prime}\mp\l^\mp C^{0\prime})\n_\pm
+\dot\w_\pm\pm\l^\mp\w'_\pm\mp{1\over2}\l^{\mp\prime}\w_\pm, \cr
& \d\L_\pm=-4\eta_{\ssW\pm}+C^\a\dl_\a\L_\pm-4C^0{}'\n_\pm
+{1\over2}(C^1{}'\pm\l^\pm C^0{}')\L_\pm \cr
& \phantom{\d\L_\pm=-}\pm{8\over\l^++\l^-}
\w_\mp\{\dot\x\pm\l^\mp\x'-(\l^+-\l^-)'
-i(\n_+\L_--\n_-\L_+)\}+4\w_\mp', \cr
& \d C^0=C^\a\dl_\a C^0+{2i\over\l^++\l^-}(\w_+^2+\w_-^2), \cr
& \d C^1=C^\a\dl_\a C^1-{2i\over\l^++\l^-}(\l^+\w_+^2-\l^-\w_-^2),\cr
& \d\w_\pm=C^\a\dl_\a\w_\pm-{1\over2}(C^{1\prime}\mp\l^\pm C^{0\prime})
\w_\mp-{2i\over\l^++\l^-}(\w_+^2+\w_-^2)\n_\pm, \cr
& \d C_\ssW=C^\a\dl_\a C_\ssW
-4i(\w_-\eta_{\ssW+}-\w_+\eta_{\ssW-}), \cr
& \d\eta_{\ssW\pm}=C^\a\dl_\a\eta_{\ssW\pm}
+{1\over2}(C^1{}'\pm\l^\pm C^0{}')\eta_{\ssW\pm} \cr
&\phantom{\d\eta_\pm=}
-{1\over\l^++\l^-}\w_\mp\{\dot C_\ssW\pm\l^\mp C_\ssW'
\mp2(\n_-\eta_{\ssW+}-\n_+\eta_{\ssW-})\} \cr
&\phantom{\d\eta_\pm=}
+{i\over2(\l^++\l^-)}\w_\pm\{(\dot\L_+-\l^+\L'_+
-{1\over2}\l^+{}'\L_+-4\n_-')\w_+ \cr
&\phantom{\d\eta_\pm{ig_{11}\over4\sqrt{-g}}\w_\pm\{}
+(\dot\L_-+\l^-\L'_-+{1\over2}\l^-{}'\L_--4\n_+')\w_-
\} \cr
&\phantom{\d\a_\pm=}
+{i\over(\l^++\l^-)^2}\w_\pm[
\{\dot\x-\l^+\x'-(\l^+-\l^-)'-i\L_+\n_-\}\n_+\w_- \cr
&\phantom{\d\a_\pm
{i\over4}\Bigl({g_{11}\over\sqrt{-g}}\Bigl)\w_\pm[}
-\{\dot\x+\l^-\x'-(\l^+-\l^-)'+i\L_-\n_+\}\n_-\w_+]. \cr
}\eqn\cbrstr
$$
 These expressions are not manifestly covariant.
It is the price of not introducing the pure gauge and auxiliary
 fields introduced in the configuration
space approach [\rHNUa]. The manifest covariance is, however, retained in
 our final expressions of the super-Weyl anomaly.

\APPENDIX{B}{B}
\centerline{\bf Nontrivial
$\W$ contains no ghost momenta
${\overline {\cal P}_\pm}~$ and ${\overline \b_\pm}$}

We shall show here the lemma in sect.4 that ghost momenta (
${\overline {\cal P}_\pm}~$ and ${\overline \b_\pm}$) dependence
 can be removed in $\W = \int d\s \omega$ up to a coboundary term,
where $\omega$ denotes the density throughout Appendix B.

We begin by noting that any operator $\omega$ with ${\rm gh}(\omega)
=2,~ {\rm dim}(\omega) = 2 c + 3$ can be expanded to linear
in ${\overline {\cal P}_+}~$ and to qubic in ${\overline \b_+}$:
$$\eqalign{
\w ={\overline {\cal P}_+}f_1 +
{\overline \b_\pm}f_2 +{\overline {\cal P}_+}{\overline \b_\pm}f_3
+\overline \b_\pm^2f_4+\overline \b_\pm^3f_5+\w_1 ,\cr
}\eqn\general
$$
where $f_i~~ (i = 1,\ldots ,5)$ and $\w_1$ are operators contatining
 no ${\overline {\cal P}_+}~$ and ${\overline \b_+}~$.
  Since $\d(\overline \b_\pm^3f_5)~$ contains a term
${\overline {\cal P}_+}\overline \b_+^2$ which can not be cancelled
with other terms unless $f_5=0$.
  Hence,
$$
f_5=0.
\eqn\flast
$$
Requiring $\d\w=\dl\chi$ for some operator $\chi~$, we obtain
$$\eqalign{
\d \w_1  =&~\vp_+f_1+{\cal J}_+f_2+\dl\chi_1 ~,\cr
\d f_1 =&{\cal C}^+f_1^\prime-{\cal C}^{+\prime} f_1-
          4i\g^{+} f_2 +{\cal J}_+f_3  ~,\cr
\d f_2 =&{\cal C}^{+}f_2^\prime-{1\over\, 2\,}{\cal C}^{+\prime}f_2
        -{1\over\, 2\,}\g^+f_1^\prime+\g^{+\prime}f_1+\vp_+f_3+
       2{\cal J}_+f_4  ~,\cr
\d f_3 =&{\cal C}^{+}f_3^\prime-{5\over\, 2\,}{\cal C}^{+\prime}f_3
        - 8i\g^+f_4 ~,\cr
\d f_4 =&~{\cal C}^{+}f_4^\prime-2{\cal C}^{+\prime} f_4
         -{1\over\, 4\,}\g^+f_3^\prime+{5\over\, 4\,}
\g^{+\prime}f_3 ~,\cr
}\eqn\fdel
$$
where
$$\eqalign{
\chi_1 =& \chi + ({\overline {\cal P}_+}{\cal C}^{+}+{1\over\, 2\,}
{\overline \b_+}\g^+)f_1-{\overline \b_+}{\cal C}^{+}f_2  \cr
        &+({\overline {\cal P}_+}{\overline \b_+}{\cal C}^{+}
+{1\over\, 4\,}\overline \b_+^2\g^+)f_3-\overline \b_+^2{\cal C}^{+}f_4.
}
$$
It can be shown that an operator
whose BRST transformation is linear in the constraints
 $\vp_+~$ and ${\cal J}_+~$ is given, up to a coboundary term
and a total divergence, by a linear combination of the constraints
and of the terms proprtinal to $Y^2_+$ and $Y_+\psi^{\prime}_+$.
  Hence, $\w_1~$ takes the form
$$
\w_1=\vp_+g_1+{\cal J}_+g_2+Y^2_+g_3+Y_+\psi_+^{\prime}g_4+\w_2,
\eqn\om
$$
where operators $g_i~~ (i = 1, \ldots,4)$ dose not depend on
${\overline {\cal P}_+},~{\overline \b_+},~Y_+~$ and $\psi_+$, and
$\w_2$ satisfies $\d\w_2=\dl\chi_2~$ for some $\chi_2$.
  Calculating $\d\w_1$ of \om\ and comparing this with the one
in \fdel, one finds that
$$\eqalign{
\d \w_2 =& \dl \chi_2, \cr
f_1 =& \d g_1 - {\cal C}^+g_1^\prime+{\cal C}^{+\prime} g_1
     +4i \g^+ g_2 - 4i\g^{+\prime}g_4 + {\cal J}_+ g_5,   \cr
f_2 =&-[~\d g_2 - {\cal C}^+g_2^\prime+{1\over\, 2\,}
{\cal C}^{+\prime} g_2
    +{1\over\, 2\,}\g^+g_1^\prime-\g^{+\prime}(g_1+4g_3)
    +{1\over\, 2\,}{\cal C}^{+\prime\prime}g_4~] \cr
     & -\vp_+g_5+{\cal J}_+ g_6,  \cr
\d g_3 =&{\cal C}^+g_3^\prime-{\cal C}^{+\prime} g_3
     +{i\over\, 2\,}\g^+g_4^\prime-{3\over\, 2\,}i\g^{+\prime}g_4,  \cr
\d g_4 =&{\cal C}^+g_4^\prime-{3\over\, 2\,}{\cal C}^{+\prime} g_4
         + 4\g^+ g_3, \cr
}\eqn\gdel
$$
where
$$\eqalign{
\chi_2 = &\chi_1 - ({\cal C}^+\vp_+ +{1\over\, 2\,}{\cal J}_+ \g_+)
 g_1 - {\cal C}^+({\cal J}_+ g_2 +Y^2_+g_3) \cr
        &-({i\over\, 2\,}\g^+Y_+ +{\cal C}^+\psi^{\prime}_+)Y_+g_4. \cr
}\eqn\cdel
$$
In \gdel\ $g_5$ and $g_6$ are any operators containing no ghost momenta
 ${\overline {\cal P}_+}$ and ${\overline \b_+}$.
They carry
 ${\rm dim}(g_5) = 3c+1/2,~{\rm dim}(g_6)=3c+1~$ and ${\rm gh}(g_5)=
{\rm gh}(g_6)=3~$.
Therefore, $g_5~$ and $g_6~$ can not contain $Y_+~$ and $\psi_+~$.
Furthermore,
\fdel\ and \gdel\ imply that
$$\eqalign{
{\cal J}_+ f_3 =&\d f_1-{\cal C}^+f_1^\prime+{\cal C}^{+\prime} f_1
          +4i\g^{+} f_2  \cr
         =&-{\cal J}_+(\d g_5-{\cal C}^+g_5^\prime+
         {5\over\, 2\,}{\cal C}^{+\prime} g_5-4i\g^+g_6)  ,\cr
{\cal J}_+ f_4 =&{1\over\, 2\,}(\d f_2-{\cal C}^+f_2^\prime+
{1\over\, 2\,}{\cal C}^{+\prime} f_2+{1\over\, 2\,}\g^+f_1^\prime
-\g^{+\prime}f_1-\vp_+f_3)  \cr
      =&-{1\over\, 2\,}{\cal J}_+ (\d g_6-{\cal C}^+g_6^\prime
  +2{\cal C}^{+\prime} g_6-{1\over\, 2\,}\g^+g_5^\prime
+{5\over\, 2\,}\g^{+\prime}g_5), \cr
}\eqn\jrel
$$
which leads to
$$
\eqalign{
f_3 =&-(\d g_5-{\cal C}^+g_5^\prime+
         {5\over\, 2\,}{\cal C}^{+\prime} g_5-4i\g^+g_6) ,\cr
f_4 =&-{1\over\, 2\,}(\d g_6-{\cal C}^+g_6^\prime
  +2{\cal C}^{+\prime} g_6-{1\over\, 2\,}\g^+g_5^\prime
+{5\over\, 2\,}\g^{+\prime}g_5). \cr
}\eqn\fsol
$$
Substituting expressions of $f_i$ ($i=1, \ldots,4)$ into \gdel\
and \fsol\ and $\w_1$ in \om\ into \general, we finally obtain
$$
\w = \w_3 + \d \eta + \dl \zeta
\eqn\omfin
$$
where
$$\eqalign{
\w_3=&\w_2 -{8\over\, 3\,}\vp_{+}g_3+{1\over\, 3\,}{\cal J}_+g^{\prime}_4
+Y^2_+g_3+Y_+\psi^{\prime}_+g_4  \cr
\eta=&-{\overline{\cal P}_+}(g_1+{8\over\, 3\,}g_3)
-{\overline \b_+}(g_2-{1\over\, 3\,}g^{\prime}_4)
+{\overline{\cal P}_+}{\overline \b_+}g_5-{1\over\, 2\,}
\overline\b_+^2g_6
  \cr
\zeta=&-({\overline{\cal P}_+}{\cal C}^++{1\over\, 2\,}
{\overline \b_+}\g^+)
(g_1+{8\over\, 3\,}g_3)+{\overline \b_+}{\cal C}^+
(g_2-{1\over\, 3\,}g^{\prime}_4)  \cr
 &+({\overline{\cal P}_+}{\overline \b_+}{\cal C}^++
{1\over\, 4\,}\overline\b_+^2\g^+)g_5
+{1\over\, 2\,}{\overline \b_+}^2{\cal C}^+g_6. \cr
}\eqn\final
$$
It is easy to show
$$
 \d\w_3=\dl\chi_3
$$
for some $\chi_3$.  The ghost momenta ${\overline{\cal P}_+}$ and
${\overline \b_+}$ are shown to appear only in the coboundary
term $\d\eta$ or in the
 total derivative term $\dl\zeta$.
  Since the same argument obviously applies to remove the dependence
 on ${\overline{\cal P}_-}$ and ${\overline \b_-}~$ from the
nontrivial solution, we have proved the lemma.

\APPENDIX{C}{C}

\centerline{\bf Derivation of the covariant form $\W_g$ of $Q^2$ anomaly}

We summarize here the derivation of $\W_g=\W+\d\X$, where $\X$ is
given in \cobaundary. The calculation will
be simplified if one uses the reflection symmetry under $\plram$. Note
that $\partial_{\sigma}$, $G_{\pm}$, $V^{\pm}_{\cal C} \rightarrow
-\partial_{\sigma}$, $-G_{\mp}$, $- V^{\mp}_{\cal C}$ while the
variables with $\x$-index are unchanged under the reflection. Using
the relations
$$
G_++G_-=2\x^\prime ,\qquad V_N^+-V_N^-=N^\x \eqn\iti
$$
and
$$\eqalign{
\d G_\pm=&{ 2\,\over {\, N^++N^-\, }}[-{\cal P}^\x +N^\mp
{\cal C}^{\x\prime}
\mp J^{+}\mp J^{-} ] \cr
}\eqn\ni
$$
with
$$\eqalign{
J^{\pm}\equiv&\pm{\, 1\,\over 2}G_\pm{\cal P}^\pm\pm{\cal P}^{\pm\prime}
         -\g_\pm^\L M^\pm+\L_\pm\b_\pm ~,\cr
}\eqn\jplamai
$$
we find that $\d\X$ is given by
$$
\d\X =k\int d\s\biggl[ \biggl( {\, 1\,\over 2}{\cal C}^\x
+V^+_{\cal C}\biggr)\d G_+
-\biggl({\, 1\,\over 4}{\cal C}^{+\prime}G_+^2
+{\cal C}^{+\prime\prime}G_+\biggr)
+L^{+}\biggr]
+\plram,
\eqn\dX
$$
where we have introduced $L^{\pm}$ by
$$\eqalign{
L^{\pm}=&-{\, i\,\over 2}(\g^\L_\pm)^2
\pm{\, i\,\over 2}[~\{\mp{\cal C}^\pm{\cal C}^{\pm\prime}
+2i(\g^\pm)^2\}\L_\pm\L_\pm^\prime
-{\cal C}^\pm (\g^\L_\pm\L^\prime_\pm-\L_\pm\g^{\L\prime}_\pm)~]\cr
&+\Bigl({\, 1\,\over 2}{\cal C}^{\prime\pm}\g^\pm
-{\cal C}^\pm\g^{\prime\pm}\Bigr)
(G_\pm\L_\pm-4\L_\pm)\pm\g^\pm(G_\pm\g_\pm^\L-4\g_+^{\L\prime})\cr
&+2i(\g^\pm)^2\Bigl({\, 1\,\over 4}G_\pm^2-G_\pm^\prime\Bigr).\cr
}\eqn\eru
$$
Using \cwgh\ and \ni, we obtain
$$\eqalign{
\d\X =k\int d\s\biggl[ &{ 1\over {\, N^++N^-\,}}C_\ssW\{-{\cal P}^\x
+2J^{+}+N^+{\cal C}^{\x\prime}\}
+V_{\cal C}^+{\cal C}^{\x\prime} \cr
&-(V_{\cal C}^+-{\cal C}^{+\prime })
(V_{\cal C}^{+\prime}+{\cal C}^{+\prime\prime})
+K^{+}+L^{+}\biggr]+\plram,\cr
}\eqn\dXni
$$
where $K^{\pm}$ are given by
$$
K^{\pm}=(V_{\cal C}^\pm -{\cal C}^{\pm\prime})(\L_\pm\g^\pm)^\prime
-(V_{\cal C}^\pm+{\cal C}^{\pm\prime})^\prime\L_\pm\g^\prime
\pm\L_\pm\L_\pm^\prime (\g^\pm)^2.
\eqn\kpm
$$
In the integrand of the rhs of \dXni\ the tems proportional
to $C_\ssW$ can be written as
$$\eqalign{
-{\cal P}^\xi+2J^{+}+N^+{\cal C}^{\x\prime}+\plram
=&-2\dot{C}_W+(N^++N^-)C_\ssW^\prime \cr
&-\biggl[~{\, 1\,\over 2}\dl _\t (G_+-G_-) -\dl _\s(V_N^++V_N^-)~\biggr]
({\cal C}^++{\cal C}^-) \cr
&-(N^+-N^-)(V_{\cal C}^++V_{\cal C}^-)^\prime+2(Z^{+}+Z^{-}) \cr
}\eqn\jminusp
$$
with
$$\eqalign{
Z^{\pm}&=\pm[~N^\pm(\L_\pm\g^\pm)^\prime-(\L_\pm M^\pm)^\prime{\cal C}^\pm~]
\mp[~G_\pm\g^\pm M^\pm +2(\g^\pm M^\pm)^\prime~]\cr
&\hskip 1.5cm -\dl_\t(\L_\pm\g^\pm)
-(\g^\L_\pm M^\pm -\L_\pm\b^\pm)\cr
&=-2(\g^\pm W_N^\pm +M^\pm W_{\cal C}^\pm),\cr
}\eqn\z
$$
where $W_N^\pm$ and $W_{\cal C}^\pm$ are define by \wpm.
By utilizing the identities
$$
K^{\pm}+L^{\pm}=-8i(\g^{\pm\prime})^2-{\, i\,\over 2}
( W_{\cal C}^\pm)^2~,
\eqn\nokori
$$
we find that the coboundary term is finally given by
$$\eqalign{
\d\X=k\int d\s\Biggl[~
&{{{\cal C}^++{\cal C}^-}\over {N^+ +N^-}}\bigl\{ {\, 1\,\over 2}
\dl_\t( G_+-G_-) -\dl_\s( V_N^++V_N^-)\bigr\}
C_\ssW\cr
&+\biggl\{{2\over {N^++N^-}}\dl_\t C_\ssW
-{{N^+-N^-}\over {N^++N^-}}\dl_\s C_\ssW\biggr\}C_\ssW\cr
&-{2\over {N^++N^-}}(\g^+ W_N^+ +\g^- W^-_N
+M^+ W_{\cal C}^++M^- W_{\cal C}^-)
  C_\ssW \cr
&-{\, i\,\over 2}\left\{( W_{\cal C}^+)^2 +(W_{\cal C}^-)^2 \right\}\cr
&-{\cal C}^+ {\cal C}^{+\prime\prime\prime}
+{\cal C}^- {\cal C}^{-\prime\prime\prime}
-8i(\g^{+\prime})^2-8i(\g^{-\prime})^2~\Biggr]~. \cr
}\eqn\delgusai
$$
One thus obtains the result \test\ of the geometrized $\W_g$
in terms of EPS variables.

\vfill\eject
\refout
\vfill\end